\def\deg{~\textrm{deg}}

\def\nk{n_{\rm b}}

\def\Pb{P_{\rm b}}

\def\rfr#1{Equation~(\ref{#1})}
\def\rfrs#1#2{Equations~(\ref{#1})~to~(\ref{#2})}
\def\Rfr#1{Eq. (\ref{#1})}

\def\derp#1#2{\rp{\partial{#1}}{\partial{#2}}}
\def\dert#1#2{\frac{{{\textrm{d}}}{#1}}{{{\textrm{d}}}{#2}}}

\def\virg#1{``#1"}

\def\eqi{\begin{equation}}
\def\eqf{\end{equation}}
\def\eqia{\begin{eqnarray}}
\def\eqfa{\end{eqnarray}}

\def\rp#1#2{{#1\over#2}}
\def\lb#1{\label{#1}}

\def\bds#1{\boldsymbol{#1}}

\def\ton#1{\left(#1\right)}
\def\qua#1{\left[#1\right]}
\def\grf#1{\left\{#1\right\}}
\def\ang#1{\left\langle #1\right\rangle}
\documentclass[onecolumn]{aastex}

\usepackage[title]{appendix}
\usepackage{hyperref}
\usepackage{booktabs}
\usepackage[table,xcdraw]{xcolor}
\usepackage{multirow}
\usepackage{rotating,tabularx}
\usepackage{float}
\usepackage{amsmath,textgreek,w-greek,wasysym}
\usepackage[polutonikogreek,english]{babel}
\usepackage{amsthm}
\usepackage{amscd,lineno}
\usepackage{amssymb,dsfont}
\usepackage{graphicx,epsfig}
\usepackage{txfonts}
\usepackage{xr-hyper}

\RequirePackage{color}

\newcommand{\emaila}{lorenzo.iorio@libero.it}
\newcommand{\grk}[1]{\selectlanguage{polutonikogreek}
#1\selectlanguage{english}}

\allowdisplaybreaks[1]

\begin{document}

\title{Measuring the De Sitter precession with a new Earth's satellite to the $\mathbf{\simeq 10^{-5}}$ level: a proposal}

\shortauthors{L. Iorio}

\author{Lorenzo Iorio\altaffilmark{1} }
\affil{Ministero dell'Istruzione, dell'Universit\`{a} e della Ricerca
(M.I.U.R.)-Istruzione
\\ Permanent address for correspondence: Viale Unit\`{a} di Italia 68, 70125, Bari (BA),
Italy}

\email{\emaila}

\begin{abstract}
The inclination $I$ of an Earth's satellite in polar orbit undergoes a secular De Sitter precession of $-7.6$ milliarcseconds per year for a suitable choice of the initial value of its non-circulating node $\Omega$. The competing long-periodic harmonic rates of change of $I$ due to the even and odd zonal harmonics  of the geopotential vanish for either a circular or polar orbit, while no secular rates occur at all. This may open up, in principle, the possibility of measuring the geodesic precession in the weak-field limit with an accurately tracked satellite by improving the current bound of $9\times 10^{-4}$ from Lunar Laser Ranging, \textcolor{black}{which, on the other hand, may be even rather optimistic,}  by one order of magnitude, or, perhaps, even better. The most insidious competing effects are due to the solid and ocean components of the $K_1$ tide since their perturbations have nominal huge amplitudes and the same temporal pattern of the De Sitter signature. They vanish for polar orbits. Departures of $\simeq 10^{-5}-10^{-3}\deg$ from the ideal polar geometry allow to keep the $K_1$ tidal perturbations to a sufficiently small level. Most of the other gravitational and non-gravitational perturbations vanish for the proposed orbital configuration, while the non-vanishing ones either have different temporal signatures with respect to the De Sitter effect or can be modeled with sufficient accuracy. In order to meet the proposed goal, the measurement accuracy of $I$ should be better than $\simeq 35~\textrm{microarcseconds}=0.034~\textrm{milliarcseconds}$ over, say, 5 yr.
\end{abstract}

keywords{
General relativity and gravitation; Experimental studies of gravity; Experimental tests of gravitational theories; Satellite orbits; Harmonics of the gravity potential field;
}

%

\section{Introduction}
According to general relativity\footnote{For recent critical overviews of the Einsteinian theory of gravitation, see, e.g., \citet{2016Univ....2...23D} and \citet{2016arXiv160509236G}.} \citep{2015Univ....1...38I}, when a spinning gyroscope follows a geodesic trajectory in the spacetime describing the gravitational field of a static body, its spin axis, viewed in the gyro's rest frame, experiences a change in its orientation with respect to some fixed reference direction pointing to distant stars. Such a phenomenon, known as geodetic or De Sitter precession, was described for the first time by \citet{1916MNRAS..77..155D} and, later, by \citet{1918KNAB...27.214S} and \citet{1921KNAB...23..729F}. For other, more recent derivations, see, e.g.,  \citet{1970PhRvD...2.1428B,1975A&A....44..417B,1979GReGr..11..149B,1994PhRvD..49..618D}.

The geodetic precession plays a role in the binary systems hosting at least one emitting radiopulsar. Indeed, soon after the discovery of PSR B1913+16 by \citet{1975ApJ...195L..51H}, \citet{1974CRASM.279..971B} realized that studying the measured pulse shape, in particular the profile width, would allow to reveal the De Sitter effect. The first successful, although qualitative, detections were obtained partly by \citet{1989ApJ...347.1030W} and, with more confidence, by \citet{1998ApJ...509..856K} with the PSR B1913+16 system. Subsequent studies were performed by \citet{2002ApJ...576..942W}. Later, the geodetic precession was revealed also in other binary pulsars such as PSR B1534+12 \citep{2003ApJ...589..495K}, PSR J1141-6545 \citep{2005ApJ...624..906H} and PSR J1906+0746 \citep{2006ApJ...640..428L}, although with a modest accuracy; see \citet{Kramer2012} for a recent overview. The most recent and accurate measurement was performed by \citet{2008Sci...321..104B} with the double pulsar PSR J0737-3039A/B \citep{2003Natur.426..531B,2004Sci...303.1153L}; the accuracy level reached is of the order of $\simeq 13\%$.

Until now, the most accurate direct measurements of the geodetic precession have been performed in the weak-field scenario of our solar system by using both the orbital angular momentum of the Earth-Moon system as a giant gyroscope moving in the external field of the Sun
\citep{1987PhRvL..58.1062B,1988PhRvL..61.2643S,1989AdSpR...9S..75D,1991ApJ...382L.101M,1996PhRvD..53.6730W,2004PhRvL..93z1101W,2009IAU...261.0801W,2018CQGra..35c5015H} and the anthropogenic gyroscopes carried onboard the Gravity Probe B (GP-B) spacecraft orbiting the Earth \citep{2011PhRvL.106v1101E,2015CQGra..32v4001E}. While GP-B reached a relative accuracy of $3\times 10^{-3}$ \citep{2011PhRvL.106v1101E,2015CQGra..32v4001E}, the Lunar Laser Ranging (LLR) technique \citep{1994Sci...265..482D} recently allowed to obtain a measurement of such a relativistic effect accurate to about $9\times 10^{-4}$ \citep{2018CQGra..35c5015H}. \textcolor{black}{However, the actual accuracy level in such a test may be worst because of some subtle issues pertaining the treatment of certain systematic errors\footnote{\textcolor{black}{According to \citet{2018CQGra..35c5015H}, high correlations among the determined values of the parameter accounting for the geodetic precession and other geophysical and astronomical ones occurred when they were simultaneously estimated; the $9\times 10^{-4}$ uncertainty was obtained by keeping them fixed to their reference values and estimating just the relativistic parameter.}}; see Sect.~4.4 of \citet{2018CQGra..35c5015H}. In their conference proceedings, \citet{2009IAU...261.0801W} reported a relative uncertainty of the order of $4\times 10^{-3}$ from LLR, while \citet{2004PhRvL..93z1101W} reached an accuracy level of $6\times 10^{-3}$ with the same technique.}

In the present work, we show that, with a new accurately tracked Earth's satellite in circular polar orbit, it should be possible to improve the constraint by \citep{2018CQGra..35c5015H} by about one order of magnitude, or, perhaps,  even better, by measuring the De Sitter effect on the spacecraft's orbital inclination. {It is assumed that we will adopt a kinematically rotating and dynamically non-rotating \citep{1989NCimB.103...63B,1994PhRvD..49..618D} geocentric equatorial coordinate system throughout the paper. An appropriate name for the proposed satellite would, thus, be ELXIS, from \grk{<'elxis} meaning `dragging', `trailing'.}

\textcolor{black}{In \citet{2018arXiv180906119I} it is shown that, in addition to the De Sitter precession to $\simeq 10^{-5}$, also the Lense-Thirring effect \citep{1918PhyZ...19..156L} could be measured at a some percent level if an ecliptic coordinate system is used for the data analysis. Finally, it is worthwhile noticing that, at first sight, the ELXIS concept might seem nothing new with respect to the past proposal put forth by \citet{1976PhRvL..36..629V,1976CeMec..13..429V,1977JSpRo..14..474S,1978AcAau...5...77V}; Sect.~8 of \citet{2018arXiv180906119I} explains why it is not the case.}

The plan of the paper is as follows.
In Section~\ref{desi}, the De Sitter rate of change of the inclination of a test body orbiting its primary which, in turn, moves in the external gravitational field of another massive object is analytically worked out. A non-vanishing, long-term effect with a magnitude of $7.6~\textrm{mas~yr}^{-1}$ is found for the Earth-Sun scenario. Depending on the temporal behaviour of the satellite's node, it can be either a sinusoidal signal or a secular trend.
The next three Sections are devoted to the main perturbations of gravitational origin on the satellite's inclination. Section~\ref{geopot} deals with the long-term signatures induced by the even and odd zonal harmonics of the Earth's geopotential. It turns out that they all vanish if the satellite follows a circular path, or if its orbital plane is perpendicular to the Earth's equator. In  Section~\ref{marea}, the aliasing due to the Earth's solid and ocean tides is discussed. Both the solid and the ocean components of the $K_1$ tidal constituent, whose key parameters are rather poorly known at present, induce long-term rates of change on the inclination which have nominally huge amplitudes and the same temporal pattern of the De Sitter effect. Luckily, they vanish for polar orbits. The impact of deviations from such an ideal orbital configuration is discussed by finding that departures up to $\simeq 100$ times larger than those characterizing GP-B at its launch  are able to reduce the nominal tidal perturbations of $K_1$ to a sufficiently small level. The 3rd-body perturbations due to the Sun and the Moon are worked out in Section~\ref{terzo}. While the heliocentric gravitational parameter is determined with an accuracy which allows to deem  the Sun-induced effect as negligible, the lunar one is more effective in potentially impacting the satellite's inclination. However, the present-day level of accuracy of the selenocentric gravitational parameter allows to fulfil our requirements. Section~\ref{ngp} treats the non-gravitational perturbations by assuming a LAGEOS-type cannonball geodetic satellite. It turns out that none of them should pose a threat to our goals since most of them vanish for a circular polar orbit, or have temporal signatures which are distinctively different from the De Sitter one. The geomagnetic field may affect the inclination of an electrically charged satellite in a circular polar orbit with a secular trend whose residual effect, however, should be  small enough in view of the current level of accuracy in our knowledge of the Earth's magnetic dipole moment. The issue of the actual observability of a change in the inclination of the order of the De Sitter one is tackled in Section~\ref{obs}. It appears that reaching a measurement accuracy for the satellite's inclination better than $\simeq 30~\upmu\textrm{as} = 0.03~\textrm{mas}$ does not seem completely unrealistic in a near future. Section~\ref{fine} resumes our findings and offers our conclusions. A list of definitions of all the physical and orbital parameters used in the text can be found in Appendix~\ref{appena}, while the numerical values of most of them are in Appendix~\ref{appenb} along with the figures.

\section{The De Sitter orbital precessions}\lb{desi}
The perturbing De Sitter potential per unit mass \textcolor{black}{of a satellite orbiting the Earth which, in turns, moves in the external field of the Sun} is \citep{1970PhRvD...2.1428B}
\eqi
U_\textrm{DS} = \rp{3\mu_\odot {\bds L}^\oplus\bds\cdot{\bds L}}{2c^2r^3_\oplus}.
\eqf
Its doubly averaged expression, obtained by using the Keplerian ellipses
as unperturbed reference orbits for both the geocentric satellite motion and the heliocentric trajectory of the Earth,
turns out to be
\eqi
\ang{U_\textrm{DS}}_{{\Pb}P_\oplus} = \rp{3\mu_\odot \nk n_\textrm{b}^\oplus a^2\sqrt{1-e^2}\qua{\cos I_\oplus\cos I + \sin I_\oplus\sin I\cos\ton{\Omega-\Omega_\oplus}}}{2c^2 a_\oplus\ton{1-e_\oplus^2}}.\lb{UdS}
\eqf
\textcolor{black}{The standard Lagrange equation for the rate of change of the inclination induced by a perturbing potential $U_\textrm{pert}$ \citep{2003ASSL..293.....B}}
\eqi
\dert{I}{t} = \rp{1}{\nk a^2 \sin I\sqrt{1 - e^2}}\ton{ \derp{U_\textrm{pert}}{\Omega} -  \cos I\derp{U_\textrm{pert}}{\omega} },\lb{dIdt}
\eqf
applied to \rfr{UdS}, allows to straightforwardly obtain the long-term, doubly averaged De Sitter rate of change of the satellite's inclination:
\eqi
\ang{\dert{I}{t}}^\textrm{DS}_{{\Pb}P_\oplus} = -\rp{3\mu_\odot n_\textrm{b}^\oplus\sin I_\oplus\sin\ton{\Omega-\Omega_\oplus}}{2c^2 a_\oplus\ton{1 - e_\oplus^2}}.\lb{IdS}
\eqf
It can be shown that \rfr{IdS} can be obtained also within the standard radial-transverse-normal  perturbative scheme by doubly averaging the right-hand-side of the Gauss equation for the variation of the inclination \citep{2003ASSL..293.....B}
\eqi
\dert{I}{t} = \rp{1}{\nk a \sqrt{1-e^2}}A_w\ton{\rp{r}{a}}\cos u
\eqf
calculated with
\begin{align}
A_w^\textrm{DS} \nonumber &= \rp{3\mu_\odot}{c^2 r_\oplus^3}\sqrt{\rp{\mu_\oplus}{a_\oplus\ton{1-e^2_\oplus}}}\qua{L_z^\oplus \sin I\ton{e\sin\omega + \sin u} + \ton{e\cos\omega + \cos u}\ton{L_x^\oplus\cos\Omega + L_y^\oplus\sin\Omega }  -\right.\\ \nonumber \\
& -\left. \cos I\ton{e\sin\omega + \sin u}\ton{ -L_y^\oplus\cos\Omega + L_x^\oplus\sin\Omega}   }.\lb{ANdS}
\end{align}
\Rfr{ANdS} is obtained by taking  the third term of Eq.~(10.12) in \citet{2010ITN....36....1P}, which describes the De Sitter acceleration, to the case of the Earth-satellite system orbiting the Sun, and projecting it onto the normal direction spanned by the out-of-plane unit vector $\hat{w}$.
The trigonometric term $\sin\ton{\Omega-\Omega_\oplus}$ entering \rfr{IdS} tells us that the De Sitter rate of change of the inclination can be viewed either as an essentially secular precession or as a long-periodic, harmonic signal depending on the frequency $\dot\Omega$ of the satellite's node and of its initial value $\Omega_0$. Indeed, given that the node of the heliocentric Earth's orbit stays constant over any conceivable time span devoted to the data analysis since its period amounts to $T_{\Omega_\oplus} = -149,229.87~\textrm{yr}$ in such a way that $\Omega_\oplus\ton{t} = \Omega_\oplus^0 + \dot\Omega_\oplus t\simeq \Omega_\oplus^0$, if the satellite's node circulates as $\Omega\ton{t} = \Omega_0 + \dot\Omega t$ and its period fulfils the condition $T_\Omega\ll T_{\Omega_\oplus}$, then the frequency of the harmonic term in \rfr{IdS} is $2\uppi\ton{1+T_\Omega T^{-1}_{\Omega_\oplus}} T^{-1}_\Omega\simeq 2\uppi T^{-1}_\Omega=\dot\Omega$.  In this case, the De Sitter effect is a harmonic one. Instead, if the satellite's node is locked in a fixed position in view of its peculiar orbital geometry which makes $\dot\Omega \simeq 0$, it is, thus, possible to obtain an essentially secular precessions for the De Sitter effect on the inclination by choosing $\Omega_0 = \Omega_\oplus + 90\deg$. From \rfr{IdS}, the magnitude of the De Sitter inclination rate is
\eqi
\ang{\dert{I}{t}}^\textrm{DS}_{{\Pb}P_\oplus} = -7.6~\textrm{mas~yr}^{-1}~\sin\ton{\Omega-\Omega_\oplus},\lb{numero}
\eqf
where $\textrm{mas~yr}^{-1}$ stands for milliarcseconds per year.

For the sake of completeness, we explicitly show also De Sitter rates of change of the satellite's node $\Omega$ and perigee $\omega$ which can be directly obtained from \rfr{UdS} with the appropriate Lagrange perturbing equations:
\begin{align}
\ang{\dert{\Omega}{t}}^\textrm{DS}_{{\Pb}P_\oplus} \lb{OdS}&= \rp{3\mu_\odot n_\textrm{b}^\oplus\qua{\cos I_\oplus -\sin I_\oplus\cot I\cos\ton{\Omega-\Omega_\oplus}}}{2c^2 a_\oplus\ton{1 - e_\oplus^2}}, \\ \nonumber \\
\ang{\dert{\omega}{t}}^\textrm{DS}_{{\Pb}P_\oplus} \lb{odS}&=  \rp{3\mu_\odot n_\textrm{b}^\oplus\sin I_\oplus\csc I\cos\ton{\Omega-\Omega_\oplus}}{2c^2 a_\oplus\ton{1 - e_\oplus^2}}.
\end{align}
It is important to note that \rfr{IdS} and \rfrs{OdS}{odS} are valid for any orbital configuration of both the satellite about its primary and of the motion of the latter one with respect to the third body.
\section{The geopotential perturbations}\lb{geopot}
A major source of systematic bias is represented, in principle, by the competing long-term classical orbital variations induced by the even and odd zonal multipoles in terms of which the departures from spherical symmetry of the Newtonian part of the Earth's gravity field are expressed \citep{HM67,Kaula00}. In particular, the node and the perigee of an Earth's satellite undergo, among other things, secular precessions due to the even zonal harmonics $J_{\ell},~\ell=2,~4,~6,\ldots$ of the geopotential \citep{HM67,Kaula00}. As such, their mismodeled components pose a major threat to a clean measurement of the relativistic signatures of interest depending on the level of uncertainty in our knowledge of $J_\ell$.

On the other hand, the satellite's inclination does not suffer from such an important drawback, as we will show below.
Here, we look at the orbital motion of a spacecraft around the Earth, assumed non spherically symmetric, and analytically calculate the rates of change of $I$ averaged over one full orbital period $\Pb$ induced by the first five zonal harmonics $J_{\ell}$ of the geopotential.
To this aim, we use \rfr{dIdt}
where the correction of degree $\ell$ to the Newtonian monopole
\eqi
U_{J_\ell} = \rp{\mu_\oplus}{r}J_{\ell}\ton{\rp{R_\oplus}{r}}^\ell \mathcal{P}_\ell\ton{\bds{\hat{r}}\bds\cdot{\bds{\hat{S}}}_\oplus},
\eqf
which replaces $U_\textrm{pert}$,
is straightforwardly averaged over one full orbital revolution by using the Keplerian ellipse
%
as reference unperturbed orbit.
As a result, no secular precessions occur for the inclination. Indeed, only long-periodic effects having harmonic patterns characterized by integer multiples of the frequency of the perigee motion are obtained. They turns out to be
\begin{align}
\ang{\dert{I}{t}}_{\Pb}^{J_2} \lb{IJ2} & = 0, \\ \nonumber \\
\ang{\dert{I}{t}}_{\Pb}^{J_3} \lb{IJ3}& = \rp{3 J_3 e \nk R_\oplus^3 \cos I \ton{3 + 5 \cos 2I} \cos\omega }{
 16 a^3 \ton{1 - e^2}^3}, \\ \nonumber \\
\ang{\dert{I}{t}}_{\Pb}^{J_4}& = \rp{15 J_4 e^2 \nk R_\oplus^4 \ton{5 + 7 \cos 2I} \sin 2I \sin 2\omega }{128 a^4 \ton{1 - e^2}^4}, \\ \nonumber \\
\ang{\dert{I}{t}}_{\Pb}^{J_5} \nonumber & = -\rp{15 J_5 e \nk R_\oplus^5} {2048 a^5 \ton{1 - e^2}^5} \qua{\ton{4 + 3 e^2}\ton{58 \cos I + 49 \cos 3 I + 21 \cos 5 I} \cos\omega + \right.\\ \nonumber \\
& + \left. 14 e^2 \ton{23\cos I + 9 \cos 3I}  \sin^2 I\cos 3\omega}, \\ \nonumber \\
\ang{\dert{I}{t}}_{\Pb}^{J_6} \nonumber & = \rp{105 J_6 e^2 \nk R_\oplus^6}{32768 a^6 \ton{1 - e^2}^6} \qua{-5 (2 + e^2) \ton{37 \sin 2I + 60 \sin 4I +
33 \sin 6I} \sin 2\omega - \right. \\ \nonumber \\
&-\left. 24 e^2 \ton{29 \cos I + 11 \cos 3I} \sin^3 I \sin 4\omega}.\lb{IJ6}
\end{align}
In the calculation, the Earth's symmetry axis ${\bds{\hat{S}}}_\oplus$ was assumed to be aligned with the reference $z$ axis; moreover, no a-priori simplifying assumptions concerning the orbital geometry of the satellite were made.
It is important to note that the largest zonal harmonic, i.e. $J_2$, does not contribute at all to the long-term variation of $I$, as per \rfr{IJ2}. Moreover, \rfrs{IJ3}{IJ6} vanish for either circular ($e=0$) or polar ($I = 90\deg$) orbits.
%
\section{The solid and ocean tidal perturbations}\lb{marea}
A further class of competing long-term gravitational orbital perturbations is represented by the solid and ocean tides \citep{2001CeMDA..79..201I,2002CeMDA..82..301K}.

Among them, the tesseral ($m=1$) $K_1$ tide, with Doodson number (165.555), is the most insidious one since it induces, among other things, long-periodic, harmonic orbital perturbations having large nominal amplitudes and the same frequency of the satellite's node. In the case of the inclination, the largest contribution to the long-term rate of change of the inclination induced by both the solid and the ocean components of $K_1$ ($\ell=2,~m=1,~p=1,~q=0$) is proportional to
\eqi
\ang{\dert{I}{t}}^{K_1}_{\Pb} \propto \rp{\cos I}{\nk a^5\ton{1-e^2}^2};\lb{tideI}
\eqf
it vanishes for strictly polar orbits.
The complete expressions for the tidal rates of change of $I$ can be obtained by applying \rfr{dIdt} to Eq.~(18) and Eq.~(46) of \citet{2001CeMDA..79..201I} with the minus sign because of a different sign convention for $U_\textrm{pert}$ adopted there; they are
\begin{align}
\ang{\dert{I}{t}}^\textrm{solid}_{\Pb} = -\sqrt{\rp{5}{24\uppi}}\rp{3g_\oplus R_\oplus ^3k_{2,1,K_1}^{\ton{0}}H_2^1\ton{K_1}\cos I}{2\nk a^5\ton{1-e^2}^2}\sin\ton{\Omega - \delta_{2,1,K_1}}\lb{solK1}, \\ \nonumber \\
\ang{\dert{I}{t}}^\textrm{ocean}_{\Pb} = \rp{6 G\rho_\textrm{w} R_\oplus ^4 \ton{1+k_2^{'}}C^{+}_{2,1,K_1}\cos I}{5\nk a^5\ton{1-e^2}^2}\cos\ton{\Omega - \varepsilon^{+}_{2,1,K_1}}.\lb{ocK1}
\end{align}
Note that, for an a-priori established satellite's node rate $\widetilde{\dot\Omega}$, which is largely determined by the first even zonal harmonic according to
\eqi
\widetilde{\dot\Omega} \simeq -\rp{3}{2}\nk\ton{\rp{R}{a}}^2\rp{J_2 \cos I }{\ton{1-e^2}^2},\lb{ssync}
\eqf
\rfr{tideI} is nearly independent of the semimajor axis $a$.

\begin{figure*}
\centerline{
\vbox{
\begin{tabular}{cc}
\epsfysize= 5.0 cm\epsfbox{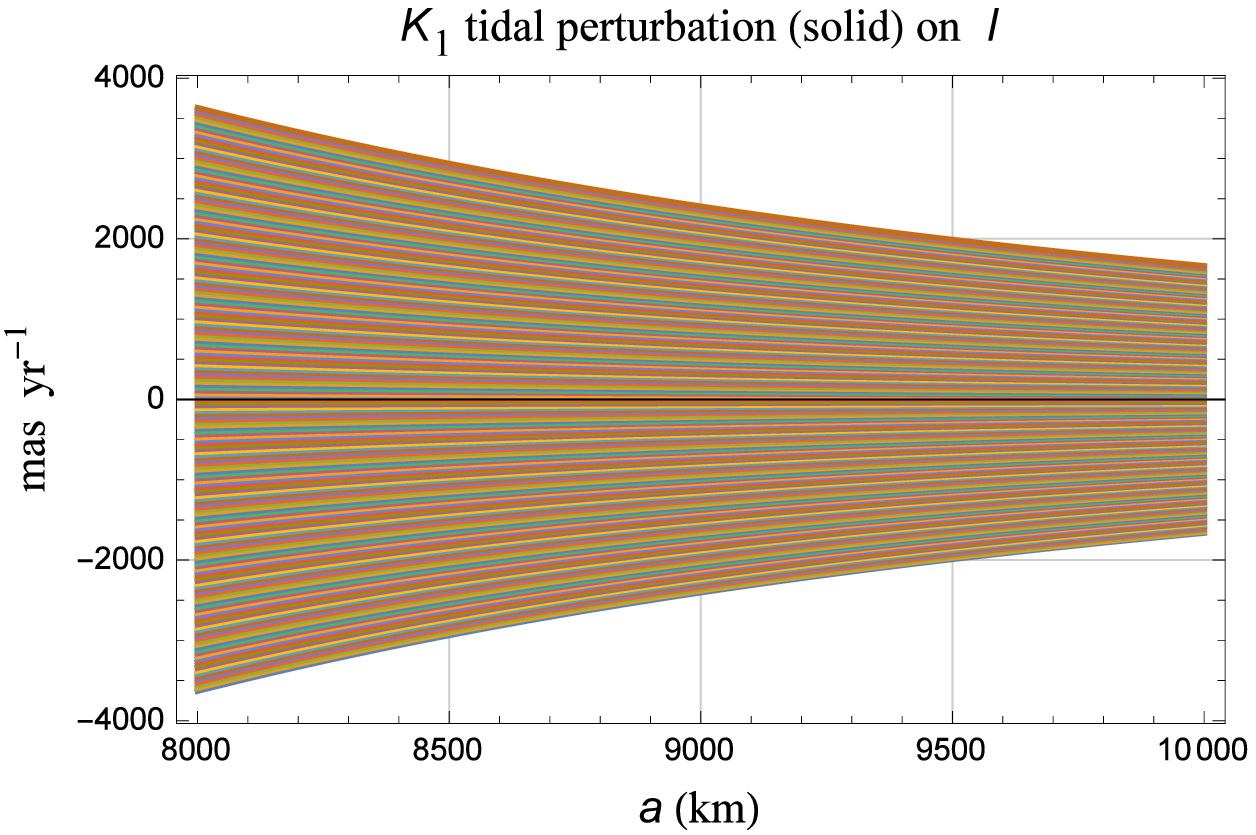} & \epsfysize= 5.0 cm\epsfbox{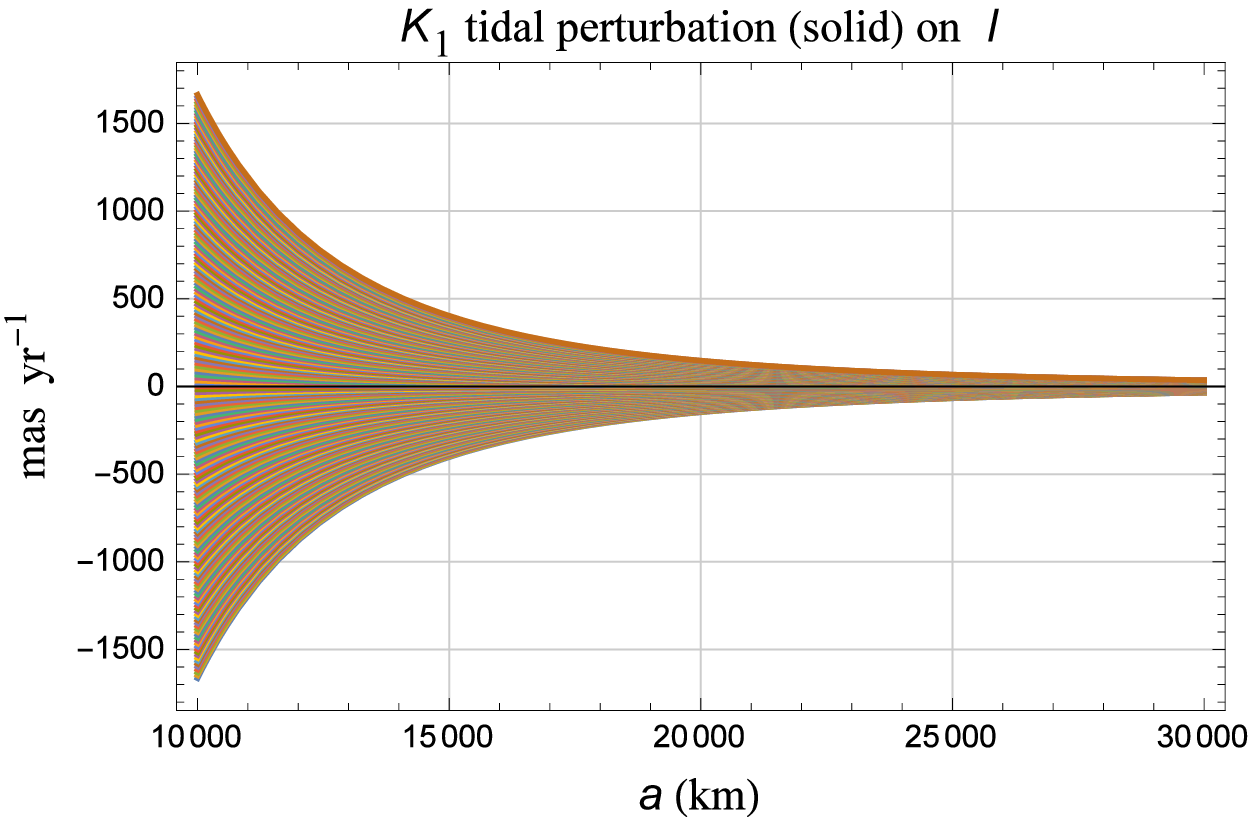}\\
\epsfysize= 5.0 cm\epsfbox{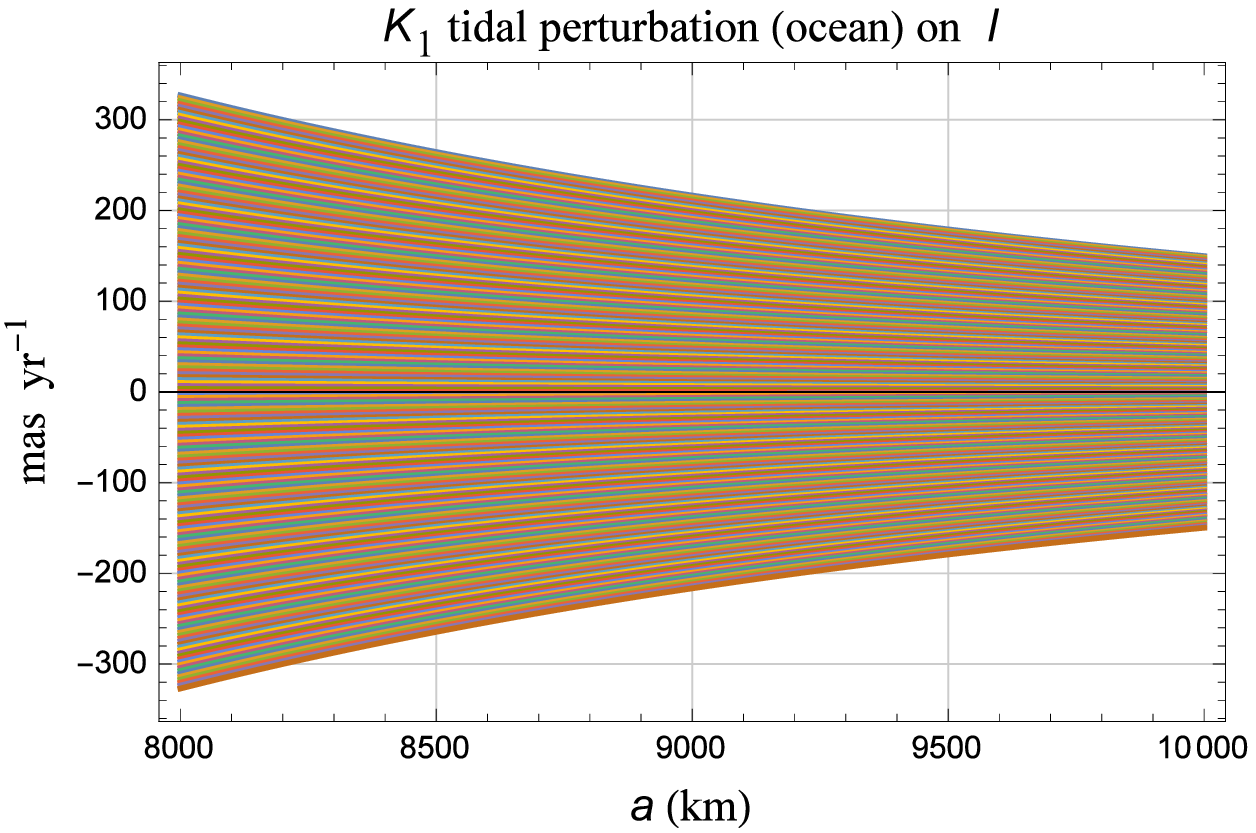} & \epsfysize= 5.0 cm\epsfbox{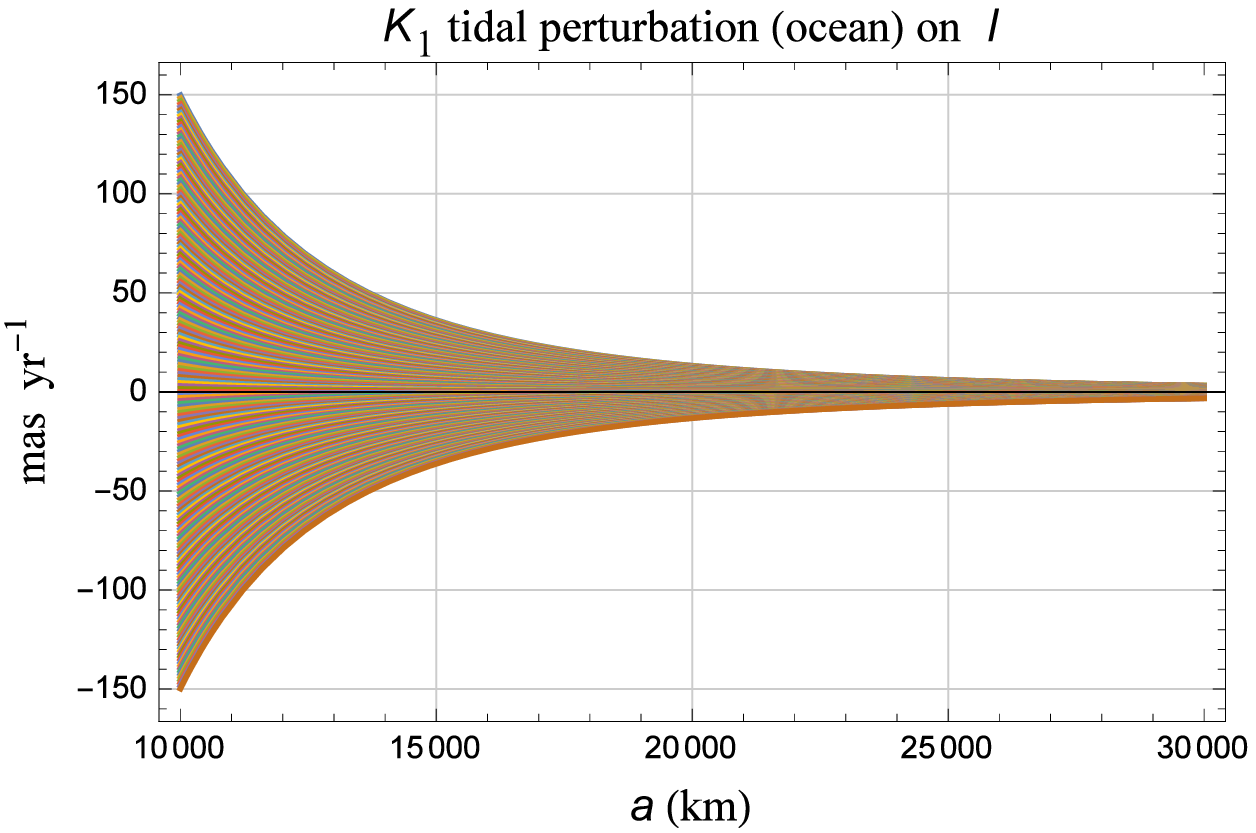}\\
\end{tabular}
}
}
\caption{Nominal amplitudes, in $\textrm{mas~yr}^{-1}$, of the rates of change of the satellite's inclination $I$ induced by the solid (upper row) and ocean prograde (lower row) components of the $K_1$ tide for $\ell=2,~m=1,~p=1,~q=0$ from \rfrs{solK1}{ocK1} as a function of the semimajor axis $a$ for different values of $I$ in the range $80\deg\leq I\leq 100\deg$. The current levels of mismodeling in $k^{(0)}_{2,1,{K_1}},~C^{+}_{2,1,{K_1}}$ are about $\simeq 10^{-3}$ \citep{2001CeMDA..79..201I} \textcolor{black}{or $3\times 10^{-4}$ \citep{polacchi018}, and $4\times 10^{-2}$ \citep{EGM96} or, perhaps, even better ($\simeq 10^{-3}$) if the global ocean models TPXO.6.2 \citep{2002JAtOT..19..183E}, GOT99 \citep{got99} and FES2004 \citep{2006OcDyn..56..394L} are compared, respectively.} }\label{tidK1}
\end{figure*}

The largest effect comes from the solid component, whose rate of change is proportional to $\sin\ton{\Omega - \delta_{2,1,K_1}}$, as per \rfr{solK1}; see the upper row of Figure~\ref{tidK1} for a plot of the nominal amplitudes of the rate of change of $I$ as a function of $a$ for different values of the inclination within the broad range $80\deg\leq I\leq 100\deg$. The uncertainty in the Love number of degree $\ell=2$ and order $m=1$ entering the amplitude of the $K_1$-induced perturbation should still be of the order of\footnote{L. Petrov and R. Ray, personal communications, August 2018.} $\simeq 10^{-3}$ \citep{2001CeMDA..79..201I}; \textcolor{black}{however, a recent data analysis of long data records of the existing LAGEOS and LAGEOS II satellites by \citet{polacchi018} reported a determination of a generic Love number $k_2$ accurate to the $3\times 10^{-4}$ level.} The ocean prograde perturbation, proportional to $\cos\ton{\Omega-\varepsilon_{2,1,K_1}^{+}}$ according to \rfr{ocK1}, has a smaller amplitude, as shown by the lower row of Figure~\ref{tidK1}.
\textcolor{black}{On the other hand, the relative mismodeling in the $C_{2,1,K_1}^{+}$ ocean tidal height coefficient  entering \rfr{ocK1} is  $4\times 10^{-2}$ \citep{EGM96}, or even at the $\simeq 10^{-3}$ level if some more recent global ocean models like TPXO.6.2 \citep{2002JAtOT..19..183E}, GOT99 \citep{got99} and  FES2004 \citep{2006OcDyn..56..394L} are compared each other.}

\begin{figure*}
\centerline{
\vbox{
\begin{tabular}{cc}
\epsfysize= 5.0 cm\epsfbox{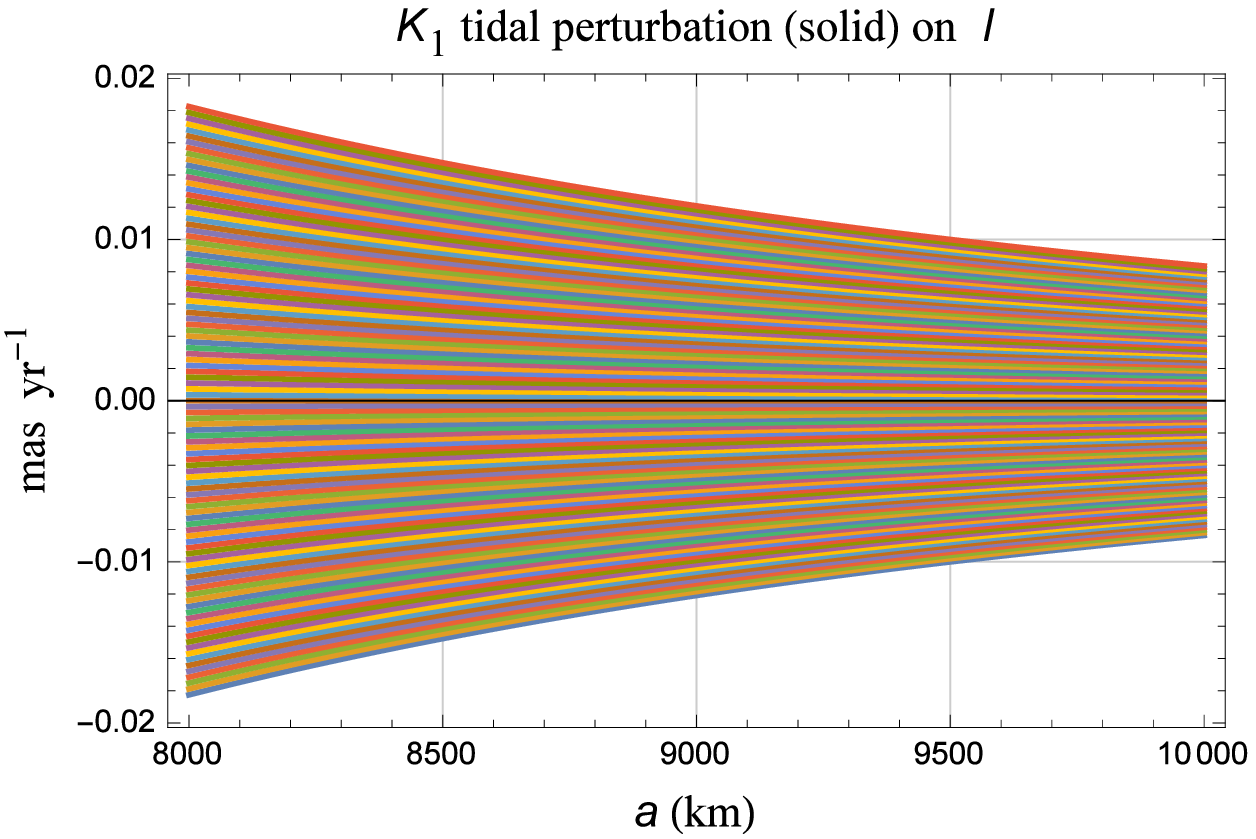} & \epsfysize= 5.0 cm\epsfbox{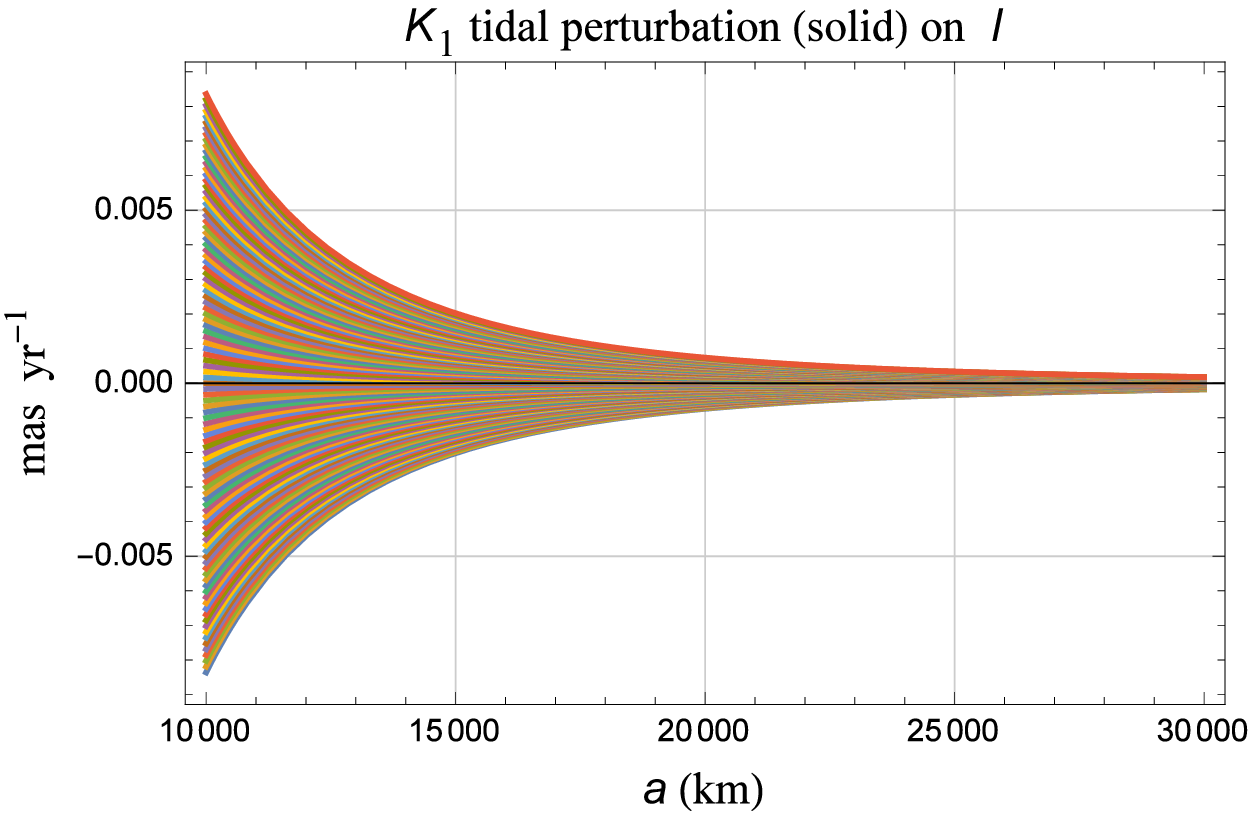}\\
\epsfysize= 5.0 cm\epsfbox{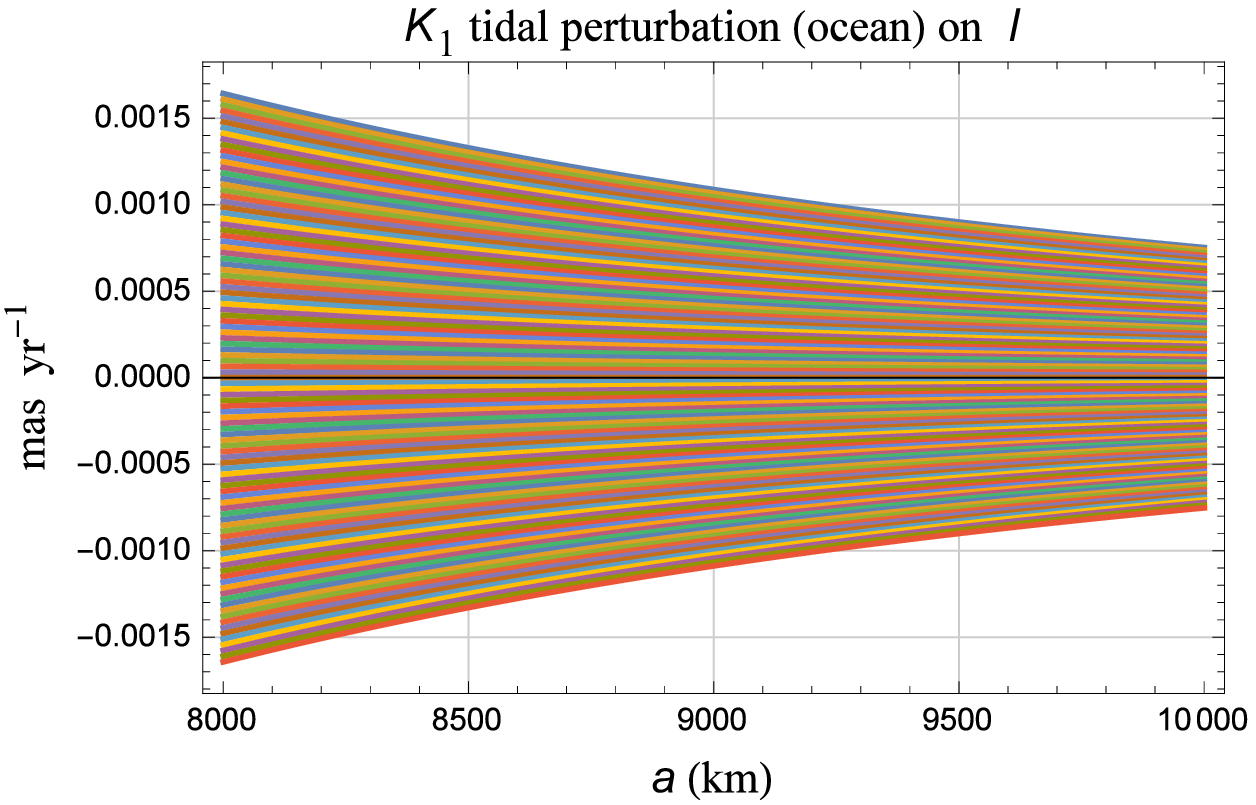} & \epsfysize= 5.0 cm\epsfbox{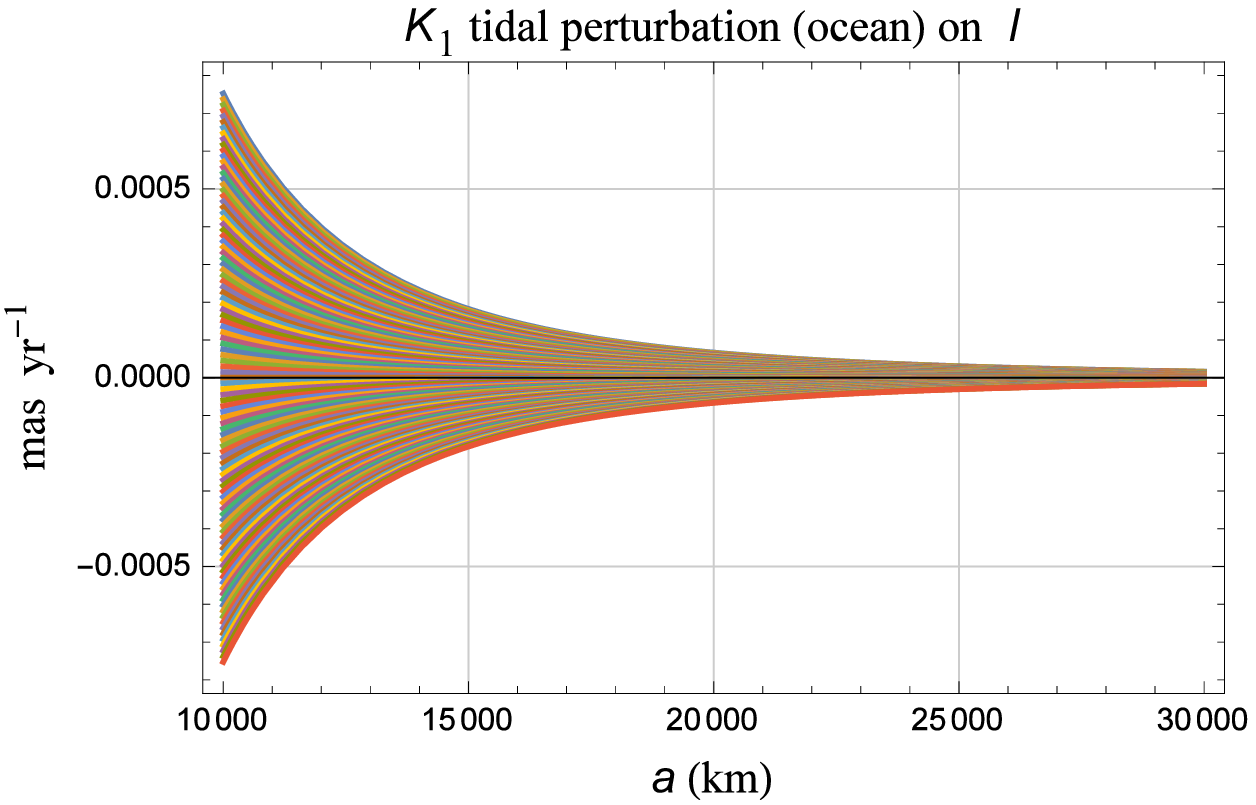}\\
\end{tabular}
}
}
\caption{Nominal amplitudes, in $\textrm{mas~yr}^{-1}$, of the rates of change of the satellite's inclination $I$ induced by the solid (upper row) and ocean prograde (lower row) components of the $K_1$ tide for $\ell=2,~m=1,~p=1,~q=0$ from \rfrs{solK1}{ocK1} as a function of the semimajor axis $a$ for different values of $I$ in the same range $I=90\pm 5\times 10^{-5}\deg$ of GP-B at its launch \citep[p. 141]{gpbrep}. The current levels of mismodeling in $k^{(0)}_{2,1,{K_1}},~C^{+}_{2,1,{K_1}}$ are about $\simeq 10^{-3}$ \citep{2001CeMDA..79..201I} or \textcolor{black}{$3\times 10^{-4}$ \citep{polacchi018}, and $4\times 10^{-2}$ \citep{EGM96} or, perhaps, even better ($\simeq 10^{-3}$) if the global ocean models TPXO.6.2 \citep{2002JAtOT..19..183E}, GOT99 \citep{got99} and  FES2004 \citep{2006OcDyn..56..394L} are compared, respectively.}}\label{gpbtid}
\end{figure*}

\begin{figure*}
\centerline{
\vbox{
\begin{tabular}{cc}
\epsfysize= 5.0 cm\epsfbox{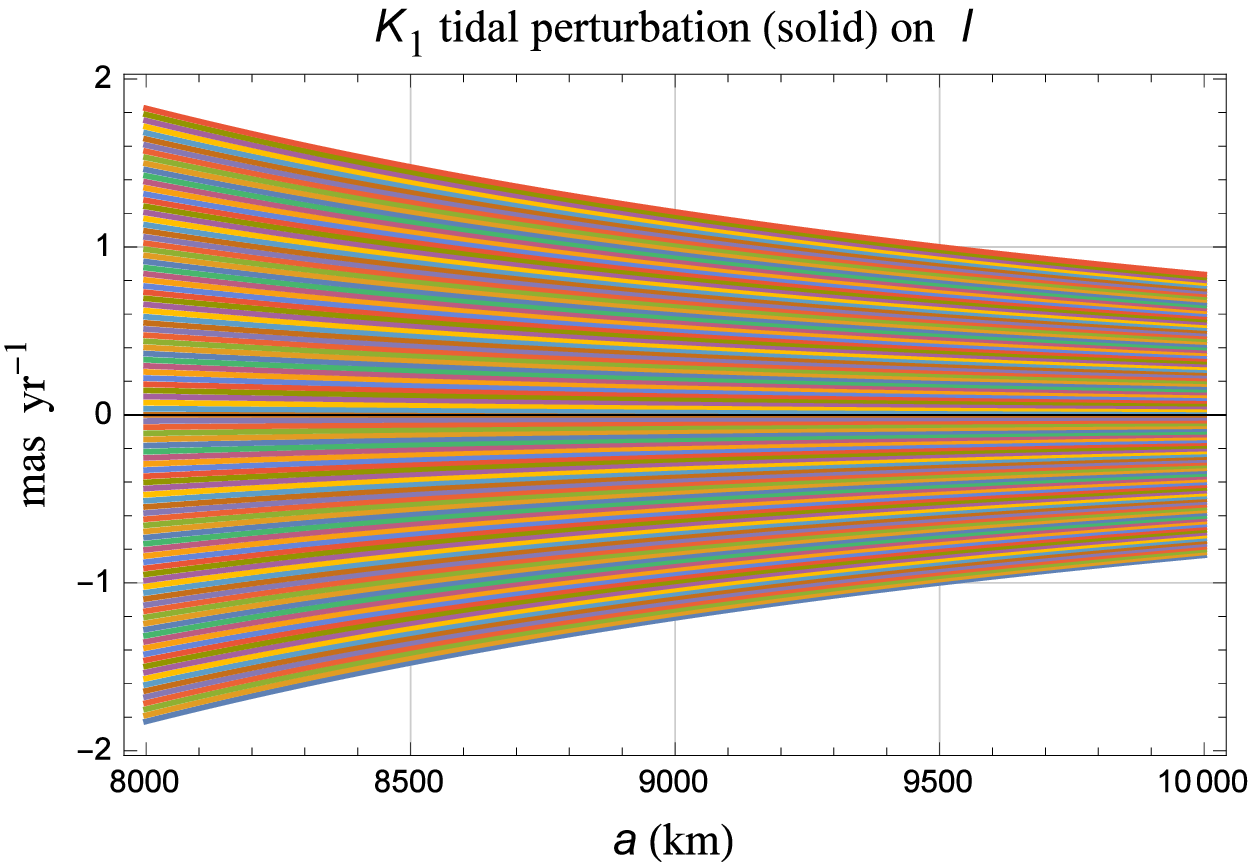} & \epsfysize= 5.0 cm\epsfbox{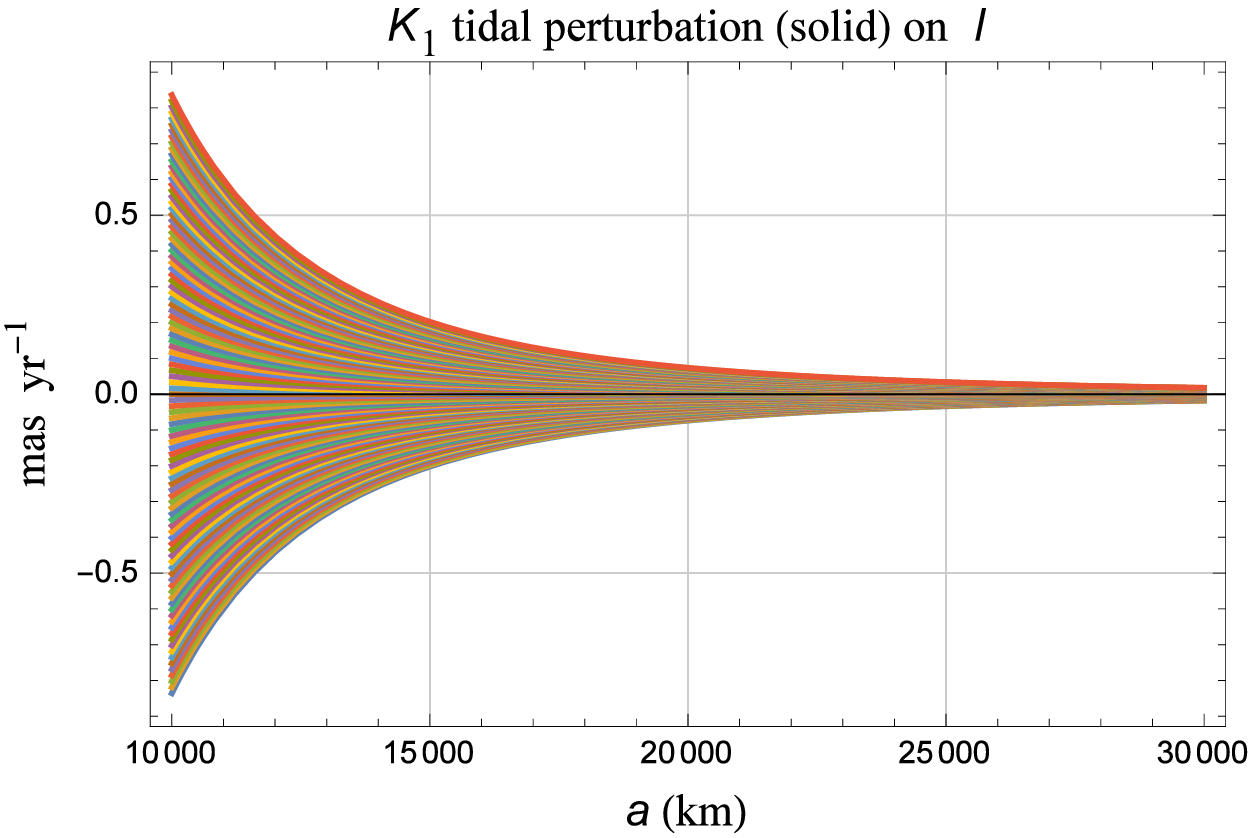}\\
\epsfysize= 5.0 cm\epsfbox{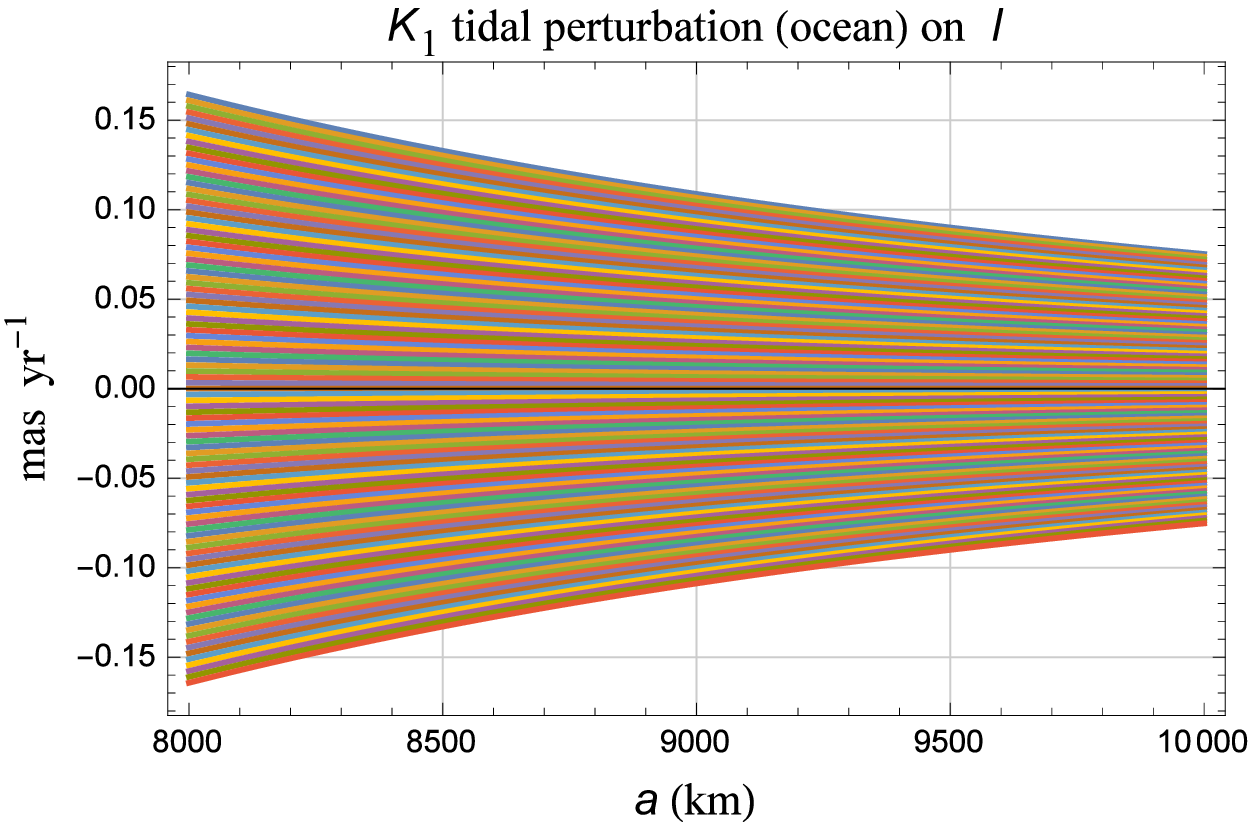} & \epsfysize= 5.0 cm\epsfbox{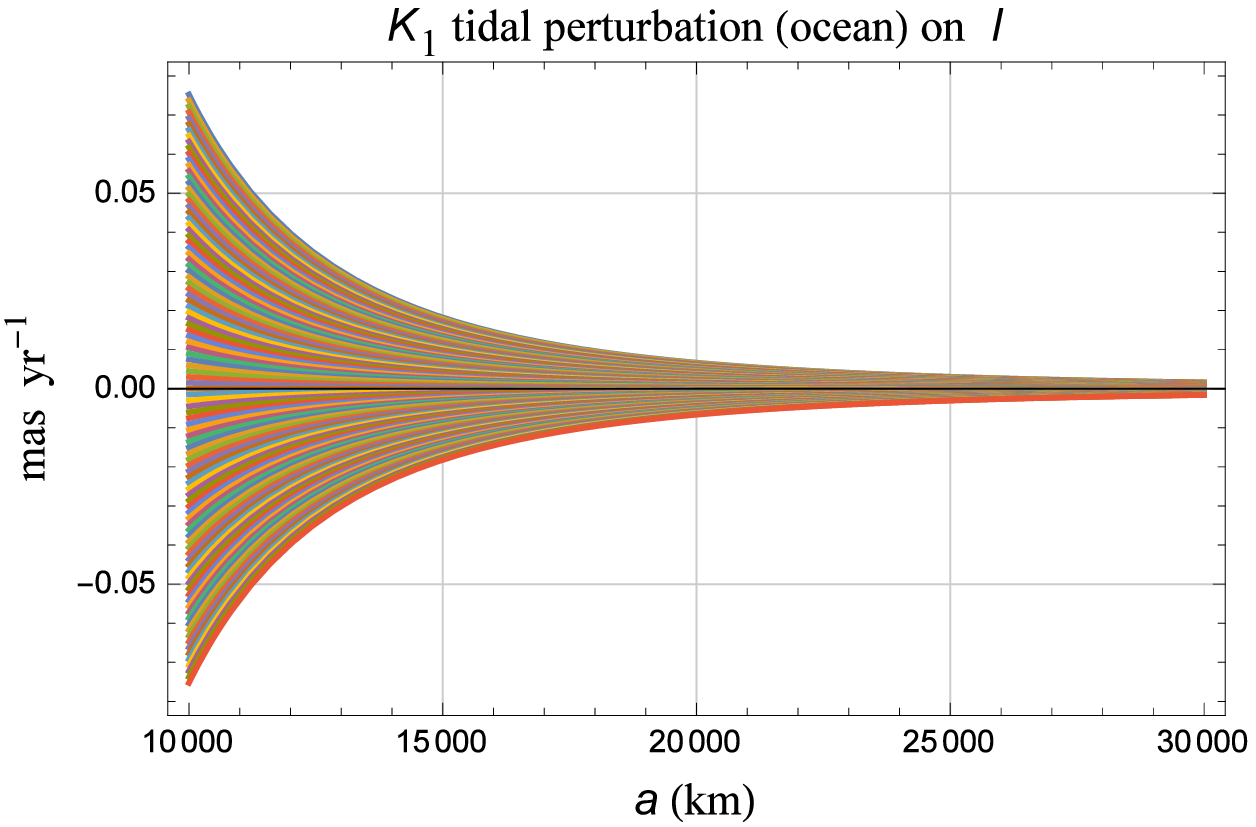}\\
\end{tabular}
}
}
\caption{Nominal amplitudes, in $\textrm{mas~yr}^{-1}$, of the rates of change of the satellite's inclination $I$ induced by the solid (upper row) and ocean prograde (lower row) components of the $K_1$ tide for $\ell=2,~m=1,~p=1,~q=0$ from \rfrs{solK1}{ocK1} as a function of the semimajor axis $a$ for different values of $I$ in the range $I = 90\pm 5\times 10^{-3}\deg$. The current levels of mismodeling in $k^{(0)}_{2,1,{K_1}},~C^{+}_{2,1,{K_1}}$ are about $\simeq 10^{-3}$ \citep{2001CeMDA..79..201I} or \textcolor{black}{$3\times 10^{-4}$ \citep{polacchi018}, and $4\times 10^{-2}$ \citep{EGM96} or, perhaps, even better ($\simeq 10^{-3}$) if the global ocean models TPXO.6.2 \citep{2002JAtOT..19..183E}, GOT99 \citep{got99} and  FES2004 \citep{2006OcDyn..56..394L} are compared, respectively.}}\label{relaxtid}
\end{figure*}

A strict polar orbital configuration can bring the nominal $K_1$ tidal perturbations significantly below  \rfr{IdS}, so that their currently assumed mismodeling, or even worse, is quite able to fulfil our requirements. Figure~\ref{gpbtid} shows the case of a circular polar orbit with the same departures from the ideal polar geometry of GP-B at its launch \citep[p.~141]{gpbrep}, i.e. $I = 90\pm 5\times 10^{-5}\deg$. However, also less tight constraints on $I$ may be adequate for our goals, especially if orbits with $a \gtrsim 10,000~\textrm{km}$ are considered. Figure~\ref{relaxtid} depicts a scenario for a circular and nearly polar orbit with $I = 90\pm 5\times 10^{-3}\deg$.

\textcolor{black}{We note that, for a fixed node orbital configuration with $\Omega\simeq \Omega_0 = \Omega_\oplus + 90~\textrm{deg}\simeq 450~\textrm{deg}$, the node-dependent trigonometric functions entering \rfrs{solK1}{ocK1} reduce to $\cos\delta_{2,1,K_1} =0.955,~\sin\varepsilon^{+}_{2,1,K_1} = -0.635$, respectively, thus further improving the overall tidal error budget. Indeed, it should be recalled that the total tidal rates of change are obtained by scaling the amplitudes plotted in Figures~\ref{tidK1}~to~\ref{relaxtid} by the aforementioned trigonometric functions of the solid and ocean tidal lag angles.}
\section{The 3rd-body perturbations: the Sun and the Moon}\lb{terzo}
Another source of potential systematic uncertainty of gravitational origin is represented by the 3rd-body perturbations induced by a distant mass X. Its doubly averaged effect on the satellite's inclination can be worked out by averaging Eq.~(7) of \citet{2012CeMDA.112..117I} over the orbital period $P_\textrm{X}$ of X. The general result is
\begin{align}
\ang{\dert{I}{t}}^\textrm{X}_{{\Pb}P_\textrm{X}} \nonumber & = \rp{3Gm_\textrm{X}}{8\nk\sqrt{1-e^2}a^3_\textrm{X}\ton{1-e_\textrm{X}}^{3/2}}\qua{\cos I \cos I_\textrm{X} + \sin I \sin I_\textrm{X}\cos\ton{\Omega - \Omega_\textrm{X}} }\times\\ \nonumber \\
\nonumber &\times \grf{5 e^2 \qua{-\sin I\cos I_\textrm{X}  + \cos I  \sin I_\textrm{X}\cos\ton{\Omega-\Omega_\textrm{X}}} \sin 2\omega + \right.\\ \nonumber \\
&+\left. \ton{2 + 3 e^2 + 5 e^2\cos 2\omega} \sin I_\textrm{X} \sin\ton{\Omega-\Omega_\textrm{X}}}.\lb{dIdtX}
\end{align}
For $e=0,~I=90\deg$, \rfr{dIdtX} reduces to
\eqi
\ang{\dert{I}{t}}^\textrm{X}_{{\Pb}P_\textrm{X}} = \rp{3Gm_\textrm{X}\sin^2 I_\textrm{X}\sin 2\ton{\Omega-\Omega_\textrm{X}}}{8\nk a^3_\textrm{X}\ton{1-e_\textrm{X}}^{3/2}}.\lb{thirdpolar}
\eqf
For a terrestrial satellite, the most important contributions to \rfrs{dIdtX}{thirdpolar} are due to the Moon and the Sun.

The heliocentric gravitational parameter $\mu_\odot$ is known with a relative accuracy of $7\times 10^{-11}$ \citep{2015JPCRD..44c1210P}; since the nominal value of \rfr{dIdtX} varies within $\simeq 10^4-10^5~\textrm{mas~yr}^{-1}$ for a satellite's circular polar orbit with $a$ ranging from, say, $10,000$ km to $30,000$ km, the systematic bias due to the 3rd-body solar perturbation can be deemed as negligible with respect to \rfr{IdS}; the same holds, a fortiori, for lower altitudes.

\begin{figure*}
\centerline{
\vbox{
\begin{tabular}{cc}
\epsfysize= 5.0 cm\epsfbox{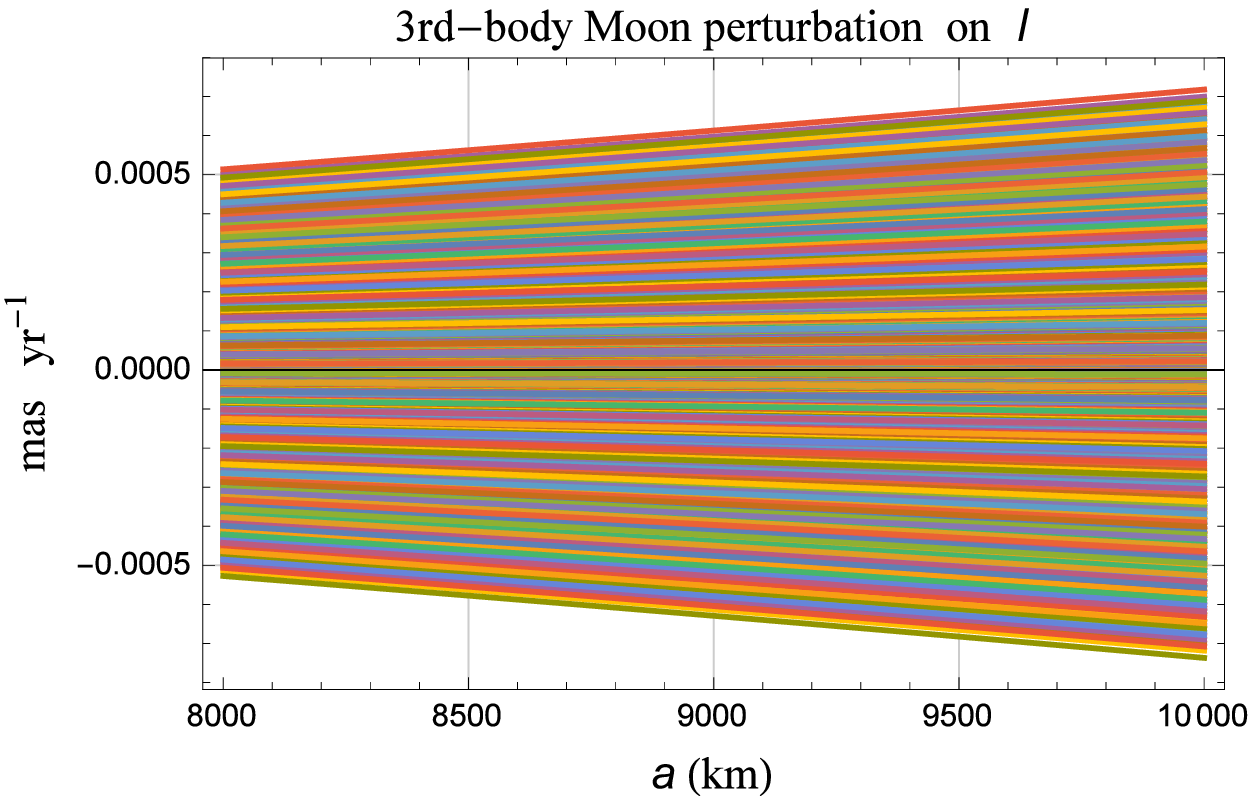} & \epsfysize= 5.0 cm\epsfbox{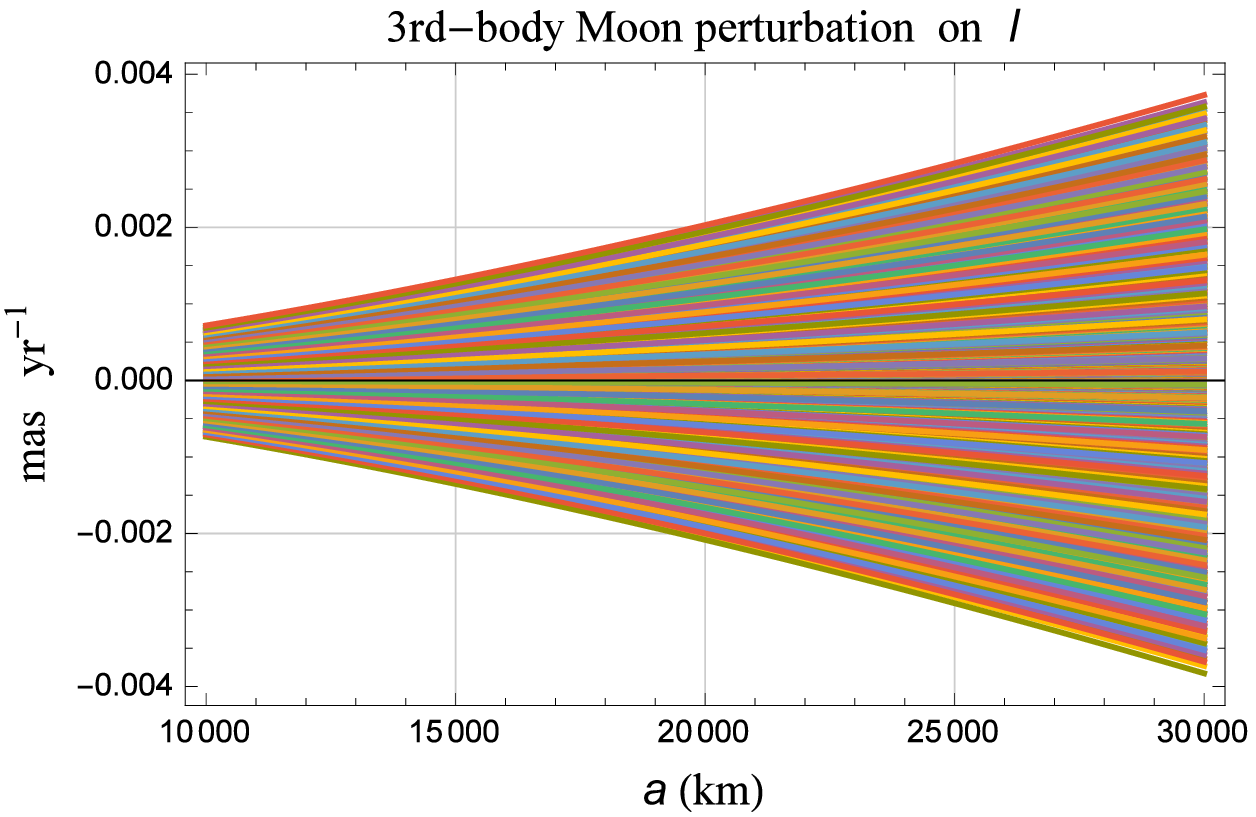}\\
\end{tabular}
}
}
\caption{
Mismodeled rate of change of the satellite's inclination, in $\textrm{mas~yr}^{-1}$, due to the 3rd-body Moon perturbation as a function of the satellite's semimajor axis $a$ for $e = 0,~I = 90\deg,~\Omega = \Omega_\oplus + 90\deg$. Each curve corresponds to a given pair of values of $I_{\leftmoon},~\Omega_{\leftmoon}$ chosen within their natural range of variation $18\deg\lesssim I_{\leftmoon}\lesssim 29\deg,~ -14\deg\lesssim \Omega_{\leftmoon}\lesssim 14\deg$ \citep{Roncoli05}. A relative error of $3\times 10^{-8}$ in the selenocentric gravitational constant $\mu_{\leftmoon}$ was adopted \citep{2010ITN....36....1P}.}\label{fig2}
\end{figure*}

In the case of the Moon, the situation is subtler because of the relatively less accurate determination of its gravitational parameter $\mu_{\leftmoon}$. It should be noted that, when referred to the Earth's equator, the lunar node oscillates around zero with a period $T_{\Omega_{\leftmoon}} = 18.6~\textrm{yr}$ \citep[Fig.~(2.4)]{Roncoli05}, while the lunar inclination has a periodicity of about $20~\textrm{yr}$  \citep[Fig.~(2.4)]{Roncoli05}; thus, for a satellite with a fixed node, \rfr{thirdpolar} represents essentially a secular trend.
According to \citet{2010ITN....36....1P}, which rely upon \citet{2009CeMDA.103..365P}, the relative uncertainty $\mu_{\leftmoon}$ can be assumed of the order of\footnote{The Object Data Page of the Moon provided by the JPL HORIZONS Web interface, revised on 2013, yields a relative uncertainty in $\mu_{\leftmoon}$ of $2\times 10^{-8}$. } $3\times 10^{-8}$. It turns out that, for $e = 0,~I = 90\deg$, the variability of the Moon's inclination and node, as referred to the Earth's equator, within their natural bounds \citep{Roncoli05} $\ton{18\deg\lesssim I_{\leftmoon}\lesssim 29\deg,~ -14\deg\lesssim \Omega_{\leftmoon}\lesssim 14\deg}$  couples to the Moon's gravitational parameter uncertainty yielding a bias on \rfr{IdS} of the order of $\simeq 3\times 10^{-5}-5\times 10^{-4}$ for $a$ ranging from $8,000$ km to $30,000$ km; see Figure~\ref{fig2}. Future, likely advances in determining $\mu_{\leftmoon}$ will improve such evaluations.
%
\section{The non-gravitational perturbations}\lb{ngp}
The impact of the non-gravitational perturbations \citep{Sehnal1975,1987ahl..book.....M,1994AdSpR..14...45D}  is, in general, more difficult to be assessed because they depend, among other things, on the actual satellite's composition, shape, physical properties, rotational state. For the sake of definiteness, in the following we will consider a LAGEOS-type cannonball geodetic satellite covered by retroreflectors for Earth-based laser tracking with the Satellite Laser Ranging (SLR) technique \citep{Combrinck2010}.

As far as the direct solar radiation pressure is concerned, Eq.~(15) of  \citet{2001P&SS...49..447L} shows that, if the eclipses are neglected, the perturbation induced by it on the inclination vanish for circular orbits. If, instead, the effect of shadow is considered, non-vanishing perturbations with frequencies $\dot\Omega,~2\dot\Omega$ would occur at zero order in the eccentricity, as shown by Tab.~(5) of \citet{2001P&SS...49..447L}.

According to Eq.~(32) of \citet{2001P&SS...49..447L}, the perturbation induced by the Earth's albedo on the satellite's inclination vanishes for circular orbits if the effect of the eclipses are neglected. Instead, if the satellite enters the Earth's shadow, zero-order perturbations in $e$, some of which with frequencies $\dot\Omega$, occur \citep[p.456]{2001P&SS...49..447L}.

Eq.~(20)~to~(22) of \citet{2002P&SS...50.1067L} show that the perturbation of the satellite's inclination due to the terrestrial thermal Yarkovsky-Rubincam effect consists of three long-term components: a secular one, which vanishes for a polar orbit or if the thermal lag angle is $\theta = 0$, and two long-periodic harmonic signals with frequencies $\dot\Omega,~2\dot\Omega$  which vanish if the orientation of the satellite's spin axis $\boldsymbol{\hat{\sigma}}$ is $\sigma_z=\pm 1,~\sigma_x=\sigma_y=0$ or, for a polar orbit, if $\theta = 90\deg$.

In the case of the solar thermal Yarkovsky-Schach  effect,
Eq.~(35) of \citet{2002P&SS...50.1067L}, which includes the effect of the eclipses, tells us that, luckily, there are no long-periodic harmonic perturbations on $I$ with frequencies multiple of the satellite's nodal one.

Eq.~(43) of \citet{2002P&SS...50.1067L} tells us that the perturbation induced by a hypothetical asymmetry in the reflectivity of the satellite's surface on the inclination vanishes for circular orbits.

According to \citet[p. 176]{1981CeMec..25..169S}, the rate of change of the inclination due to the terrestrial infrared radiation pressure is proportional to $e^2$, so that it  vanishes for circular orbits.

The atmospheric drag causes a long-term variation of the satellite's inclination to the zero order in the eccentricity which, among other things, is proportional to the atmospheric density, as, e.g., per Eq.~(6.17) of  \citet{Nobilibook87}. Thus, if it experiences marked seasonal or stochastic temporal variations during the data analysis due to some physical phenomena like, e.g., the solar activity, the resulting temporal pattern may no longer be deemed as a regular trend.

\begin{figure*}
\centerline{
\vbox{
\begin{tabular}{cc}
\epsfysize= 5.0 cm\epsfbox{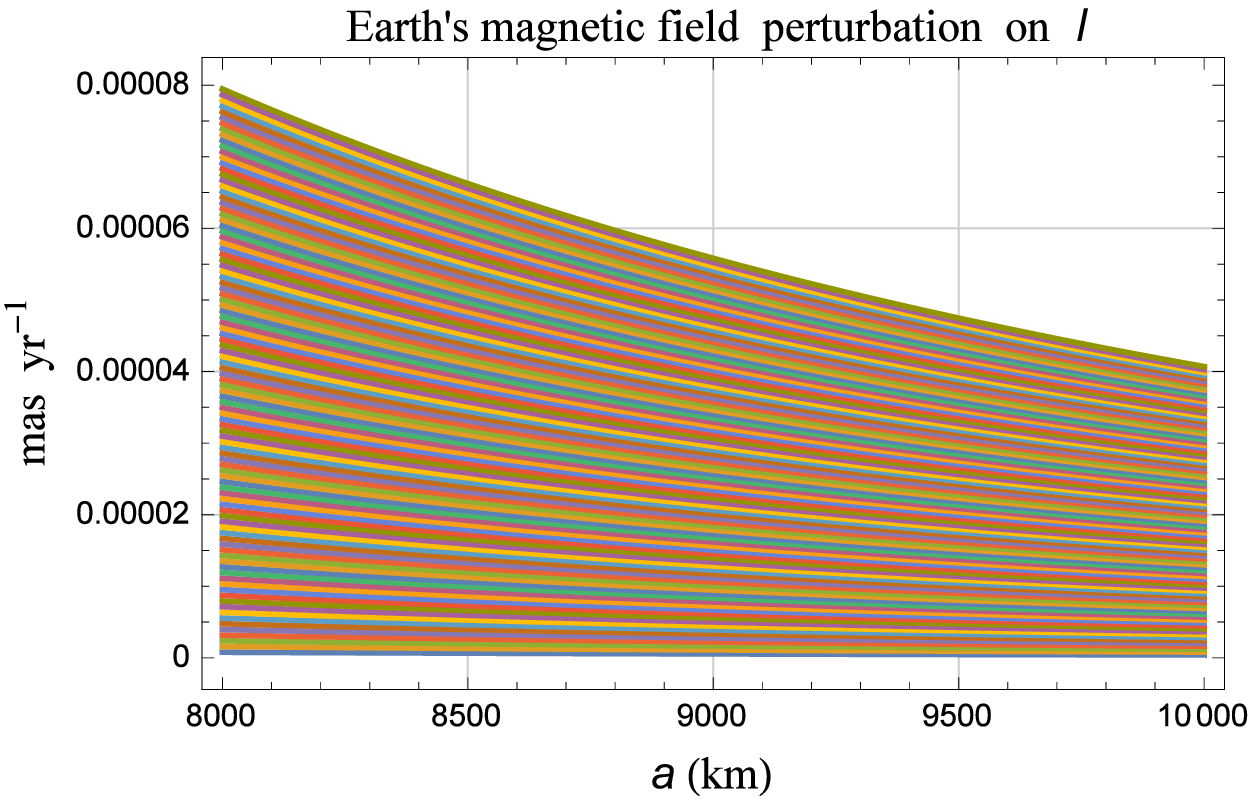} & \epsfysize= 5.0 cm\epsfbox{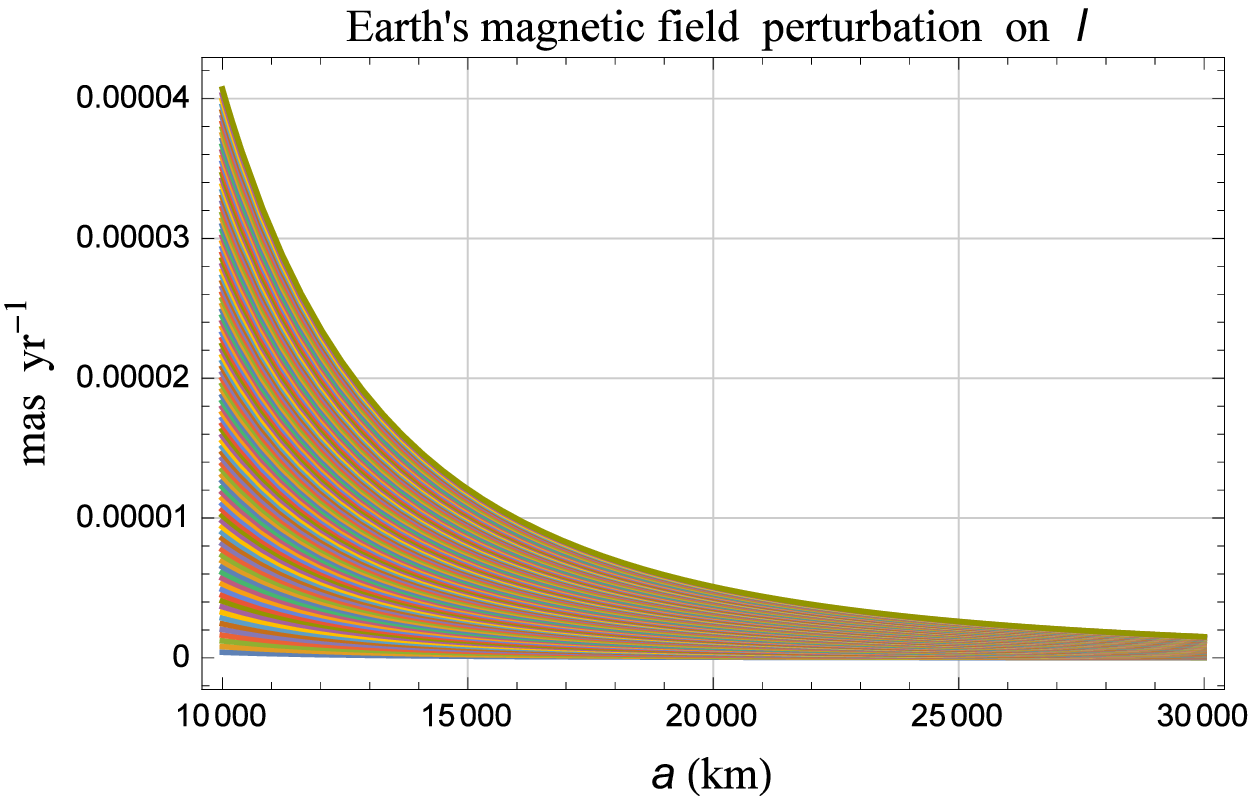}\\
\end{tabular}
}
}
\caption{Mismodeled amplitude, in $\textrm{mas~yr}^{-1}$, of the rate of change of the satellite's inclination $I$ induced by the Earth's magnetic field through the Lorentz force as a function of the semimajor axis $a$ for different values of the satellite's surface charge $\left|Q\right|$ within the range $1-100\times 10^{-11}~\textrm{C}$ admitted for LAGEOS \citep{1989CeMDA..46...85V}. A circular, polar orbit was adopted along with the mass of LAGEOS. The assumed relative uncertainty in the Earth's magnetic dipole $\textrm{m}_\oplus$ is $6\times 10^{-4}$ \citep[Tab.~1]{2009P&SS...57.1405D}. }\label{magnetico}
\end{figure*}

The interaction between the Earth's magnetic field, assumed here dipolar and with its dipole moment \textbf{m}$_\oplus$ aligned with the rotational axis, and the possible surface electric charge $Q$ of the satellite induce long-term orbital perturbations \citep{2014RAA....14..589A}. By means of Eq.~(24) in \citet{2014RAA....14..589A}, with $1/\sin f$ in its first term corrected to $\sin f$ and $B_0\rightarrow(\upmu_0/4\uppi)\textrm{m}_\oplus$,  for $e=0,~I=90\deg$ it is possible to obtain
\eqi
\ang{\dert{I}{t}}_{\Pb}^\textrm{magn} = -\rp{\upmu_0 \textrm{m}_\oplus Q}{8\uppi a^3 m_\textrm{s}}.\lb{dIdtB}
\eqf
Since the Earth's magnetic dipole is currently known with a relative accuracy of the order of $6\times 10^{-4}$ \citep[Tab.~1]{2009P&SS...57.1405D}, \rfr{dIdtB} impacts \rfr{IdS} at a $\simeq 10^{-5}$ level, as depicted by Figure~\ref{magnetico} obtained for the mass of the existing LAGEOS satellite and by varying the satellite's electric charge within $-100\times 10^{-11}~\textrm{C}\leq Q\leq -1\times 10^{-11}~\textrm{C}$ \citep{1989CeMDA..46...85V}.

Eq.~(33) of \citet{2014RAA....14..589A} shows that, for polar orbits, the inclination is not affected by electric forces of dipolar origin.

The Poynting-Robertson drag, among other things, exerts a secular drift on the inclination \citep[Eq.~(11)]{2016MNRAS.460..802L}. It  turns out to be negligible for our purposes.

As a consequence of such an analysis, it turns out that, in presence of eclipses, the solar radiation pressure and the albedo induce perturbations on $I$ having essentially the same temporal pattern of the De Sitter signal of \rfr{IdS}. Actually, as it can be inferred  from Fig.~(2) and Fig.~(3) of \citet{2015JAsGe...4..117I} and with the aid of the expressions for $\hat{w},~\hat{s}$ of \citet[p. 450]{2001P&SS...49..447L}, it turns out that, for $I = 90\deg,~\Omega = \Omega_\oplus + 90\deg$, it is not possible to avoid the entrance of the satellite into the Earth's shadow during the yearly cycle of the  solar longitude $\uplambda_\odot$ since $-1\leq \cos i_\odot\leq 1$. This would suggest to adopt a sun-synchronous orbit which, by construction,  avoids the eclipses. Indeed, in this case, the satellite's node circulates with the same period of the apparent geocentric motion of the Sun, i.e. 1 yr \citep{2014hso..book.....C}.
In order to meet such a condition, the orbital plane should be no longer polar, with an inclination depending on the adopted value of the semimajor axis. Abandoning the polar orbital configuration does not affect the previously outlined error budget, at least as far as  the static part of the geopotential is concerned, provided that the orbit is still kept circular. Indeed, \rfrs{IJ3}{IJ6} tell us that they vanish for $e=0$ independently of $I$.
If $\Omega$ were not constant, the De Sitter signature of \rfr{IdS} would look like a long-periodic, harmonic effect with the yearly period of the node. Such a choice would have the advantage of avoiding any possible competing perturbations of non-gravitational origin characterized by the same peculiar temporal pattern. Indeed, the only non-vanishing non-gravitational rates of change of $I$, i.e. the Yarkovsky-Rubincam effect and the atmospheric drag, are secular and, perhaps, stochastic or seasonal. Furthermore, a time-varying periodic signal with a definite frequency can be measured much more accurately.
On the other hand, it is an unfortunate circumstance that a sun-syncronous orbital configuration would leave very large tidal perturbations due to the solid and ocean components of the $K_1$ tide, as shown by \rfrs{solK1}{ocK1} and \rfr{ssync}. It turns out that their nominal amplitudes would amount to $4576~\textrm{mas~yr}^{-1},-412~\textrm{mas~yr}^{-1}$, respectively. The current level of mismodeling in them would not allow to meet our $\simeq 10^{-5}$ accuracy goal. Thus, a strict polar orbital configuration has to be finally deemed as preferable, although at the price of introducing potential non-gravitational effects due to the eclipses.
However, an analytical calculation of the rate of change of $I$ under the action of the direct solar radiation pressure, performed, to the zero order in $e$, by using  the first term in the series of Eq.~(2) and Eq.~(4) in \citet{1972CeMec...5...80F} for the shadow function, shows that that, for $I = 90\deg$ and $\Omega$ fixed to some given value $\widetilde{\Omega}$, no secular effects occur. Indeed, the resulting general expression is
\begin{align}
\ang{\dert{I}{t}}_{\Pb}^\textrm{srp + shadow} \nonumber &=\rp{A_\odot R_\oplus}{8\uppi\sqrt{\mu_\oplus a}}\grf{ 4 \cos  \epsilon \cos  2\Omega \sin I \sin 2\lambda_\odot - \right.\\ \nonumber \\
\nonumber &-\left. 4 \cos I \ton{\cos \Omega \sin \epsilon \sin 2\lambda_\odot + \sin 2\epsilon \sin^2\lambda_\odot \sin\Omega} - \right.\\ \nonumber \\
&-\left. \sin I \qua{\ton{3 + \cos 2\epsilon} \cos 2\lambda_\odot + 2\sin^2\epsilon}
\sin 2\Omega  }.\lb{shd}
\end{align}
For $I=90\deg$, \rfr{shd} reduces to
\eqi
\ang{\dert{I}{t}}_{\Pb}^\textrm{srp + shadow} =\rp{A_\odot R_\oplus}{8\uppi\sqrt{\mu_\oplus a}}\grf{4 \cos\epsilon \cos 2\widetilde{\Omega}
\sin 2\lambda_\odot - \qua{\ton{3 + \cos 2\epsilon} \cos 2\lambda_\odot + 2\sin^2\epsilon}\sin 2\widetilde{\Omega}},
\eqf
which  is a harmonic signal with the yearly period of the solar longitude.
The same feature holds also for the effect of the eclipses on the perturbations induced by the Earth's albedo. Indeed, they can be calculated in the same way as for the direct solar radiation pressure, apart from the modification introduced by Eq.~(36) in \citet{2001P&SS...49..447L} which does not change the frequencies of the resulting signature:
\begin{align}
\ang{\dert{I}{t}}_{\Pb}^\textrm{alb + shadow} \nonumber &=\rp{A_\textrm{alb} R_\oplus\sqrt{1 - \ton{\rp{R_\oplus}{a}}^2}}{4\uppi\sqrt{\mu_\oplus a}}\grf{ 4 \cos  \epsilon \cos  2\Omega \sin I \sin 2\lambda_\odot - \right.\\ \nonumber \\
\nonumber &-\left. 4 \cos I \ton{\cos \Omega \sin \epsilon \sin 2\lambda_\odot + \sin 2\epsilon \sin^2\lambda_\odot \sin\Omega} - \right.\\ \nonumber \\
&-\left. \sin I \qua{\ton{3 + \cos 2\epsilon} \cos 2\lambda_\odot + 2\sin^2\epsilon}
\sin 2\Omega  }.\lb{alb}
\end{align}
%
\section{Accuracy in determining the inclination}\lb{obs}
From an observational point of view, reaching the present-day LLR-based relative accuracy level of $9\times 10^{-4}$  \citep{2018CQGra..35c5015H} in measuring the shift corresponding to \rfr{IdS} over, say, 5 yr would imply an ability to determine the satellite's inclination with an accuracy of $\upsigma_I \simeq 34~\upmu\textrm{as} = 0.034~\textrm{mas}$. \citet{2009SSRv..148...71C} claimed they were able to determine the inclinations of LAGEOS and LAGEOS II, respectively, to the $\simeq 30-10~\upmu\textrm{as}$ level over $\simeq 1-3$ yr.
As far as the \virg{instantaneous} errors are concerned\footnote{K. So\'{s}nica, personal communication, August 2018.}, they are about $\simeq 10.8-18~\upmu\textrm{as}$ for 3-day solutions of GPS satellites.
The spacecraft of the Global Navigation Satellite System (GNSS)  have higher orbits than the LAGEOS' ones, and their orbits are based on continuous observations. Therefore, the angular Keplerian orbital parameters are well determined for these satellites.
Although undoubtedly challenging, it should not be, perhaps, unrealistic to expect further improvements which would allow to reach the $\simeq 10^{-5}$ level of the De Sitter effect in a foreseeable future.
\section{Summary and conclusions}\lb{fine}
The present-day best measurement of the geodetic precession has been obtained by continuously monitoring the motion of the Earth-Moon system in the field of
the Sun with the Lunar Laser ranging technique; its claimed relative accuracy is $9\times 10^{-4}$, \textcolor{black}{but it might be somewhat optimistic because of the impact of certain systematic errors. Previously published LLR-based reports yielded uncertainties of the order of $\simeq 4-6\times 10^{-3}$.}
In this paper, we showed that measuring the long-term De Sitter effect on the inclination  of a dedicated terrestrial artificial satellite to a $\simeq 1\times 10^{-4}-5\times 10^{-5}$ level should be feasible in a foreseeable future.

By adopting a circular trajectory in an orbital plane perpendicular to the Earth's equator and suitably oriented in space has several important advantages.

First, it is possible to transform the otherwise harmonic De Sitter signal having the satellite's node frequency into an essentially secular precession of $-7.6~\textrm{mas~yr}^{-1}$.

Moreover, all the competing long-term perturbations induced by the even and odd zonals of the geopotential vanish, although they have a temporal signature different from the relativistic one since their frequencies are multiple of that of the satellite's perigee.

Furthermore, also the competing long-term perturbations due to the solid and ocean components of the $K_1$ tide, which are characterized by huge nominal amplitudes and the same temporal pattern of the De Sitter signature, vanish. It is quite important since the current accuracy in knowing their key parameters is relatively modest. In order to bring their nominal signatures significantly below the threshold of the relativistic one, departures from the ideal polar configuration as little as $5\times 10^{-5}\deg$ are required, especially for relatively small values of the satellite's semimajor axis. However, even relaxing such a tight requirement by two orders of magnitude should not compromise our goal if altitudes over $3,600~\textrm{km}$ are considered.

The 3rd-body perturbations due to the Sun are far negligible since the heliocentric gravitational parameter is known with high accuracy. As far as the Moon is concerned, its impact is potentially more important; however, the present-day level of accuracy of its gravitational parameter is adequate to meet our goal for most of the satellite's altitudes considered. It is entirely plausible to assume that the continuous laser tracking of our natural satellite will further improve the determination of its gravitational parameter in the foreseeable future.

Most of the non-gravitational perturbations vanish for the orbital geometry proposed here. The remaining ones either have temporal signatures other than the De Sitter one or are modeled with a sufficiently high accuracy for our purposes.

The measurement accuracy required to improve the \textcolor{black}{allegedly optimistic $9\times 10^{-4}$ level} over, say, 5 yr is below $\simeq 30~\upmu\textrm{as}$. Depending on the actual tracking techniques which will be finally adopted, it should not be a prohibitive task to be accomplished in a not too distant future in view of the currently available results for different types of existing spacecraft.
\begin{appendices}
\section{Notations and definitions}\lb{appena}
Here, some basic notations and definitions used in the text are presented. For the numerical values of some of them, see Table~\ref{tavola0}.
\begin{description}
\item[] $G:$ Newtonian constant of gravitation
\item[] $c:$ speed of light in vacuum
\item[] $\upmu_0:$ magnetic permeability of vacuum
\item[] $M_\oplus:$ mass of the Earth
\item[] $\mu_\oplus\doteq GM_\odot:$ gravitational parameter of the Earth
\item[] ${\bds{\hat{S}}}_\oplus:$ spin axis of the Earth
\item[] $R_\oplus:$ equatorial radius of the Earth
\item[] \textbf{m}$_\oplus:$ magnetic dipole moment of the Earth
\item[] ${\overline{C}}_{\ell,m}:$ fully normalized Stokes coefficient of degree $\ell$ and order $m$ of the  multipolar expansion of the Earth's gravitational potential
\item[] $J_\ell=-\sqrt{2\ell+1}~{\overline{C}}_{\ell,0}:$ zonal harmonic coefficient of degree $\ell$ of the multipolar expansion of the Earth's gravitational potential
    \item $U_{J_\ell}:$ deviation of degree $\ell$ and order $m=0$ from spherical symmetry of the Newtonian part of the Earth's gravitational potential
\item[] $\mathcal{P}_\ell\ton{\xi}:$ Legendre polynomial of degree $\ell$
\item[] $g_\oplus:$ Earth's acceleration of gravity at the equator
\item[] $k_{2,1,K_1}^{\ton{0}}:$ dimensionless frequency-dependent Love number for the $K_1$ tidal constituent of degree $\ell=2$ and order $m=1$
\item[] $H_2^1\ton{K_1}:$ frequency-dependent solid tidal height for the $K_1$ constituent of degree $\ell=2$ and order $m=1$
\item[] $\delta_{2,1,K_1}:$ phase lag of the response of the solid Earth with respect to the constituent $K_1$ of degree $\ell=2$ and order $m=1$.
\item[] $\rho_\textrm{w}:$ volumetric ocean water density
\item[] $k_2^{'}:$ dimensionless load Love number
\item[] $C_{2,1,K_1}^{+}:$ ocean tidal height  for the constituent $K_1$ of degree $\ell=2$ and order $m=1$.
\item[] $\varepsilon^{+}_{2,1,K_1}:$ phase shift due to hydrodynamics of the oceans for the tidal constituent $K_1$ of degree $\ell=2$ and order $m=1$.
\item[] $Q:$ satellite's surface electric charge
\item[] $m_\textrm{s}:$ satellite's mass
\item[] $\boldsymbol{\hat{\sigma}}:$ satellite's spin axis
\item[] $\theta:$ satellite's thermal lag angle
\item[] $\bds r:$ satellite's position vector with respect to the Earth
\item[] $r:$ magnitude of the satellite's position vector with respect to the Earth
\item[] $\bds L:$ orbital angular momentum per unit mass of the geocentric satellite's orbit
\item[] $a:$  semimajor axis of the geocentric satellite's orbit
\item[] $\nk \doteq \sqrt{\mu_\oplus a^{-3}}:$  Keplerian mean motion of the geocentric satellite's orbit
\item[] $\Pb\doteq 2\uppi\nk^{-1}:$ orbital period of the geocentric satellite's orbit
\item[] $e:$  eccentricity of the geocentric satellite's orbit
\item[] $I:$  inclination of the orbital plane of the geocentric satellite's orbit to the Earth's equator
\item[] $\Omega:$  longitude of the ascending node  of the geocentric satellite's orbit
\item[] $\Omega_0:$  initial value of the longitude of the ascending node  of the geocentric satellite's orbit
\item[] $\dot\Omega:$ frequency of the node of the geocentric satellite's orbit
\item[] $T_{\Omega}\doteq 2\uppi \dot\Omega^{-1}:$ period of the node of the geocentric satellite's orbit
\item[] $\omega:$  argument of perigee of the geocentric satellite's orbit
\item[] $u\doteq \omega + f:$  argument of latitude of the geocentric satellite's orbit
\item[] $A_N:$ normal component of a generic satellite's perturbing acceleration
\item[] $A_\odot:$ magnitude of the satellite's disturbing acceleration due to the direct solar radiation pressure
\item[] $A_\textrm{alb}:$ magnitude of the satellite's disturbing acceleration due to the Earth's albedo
\item[] $\hat{w}=\grf{\sin I\sin\Omega,~-\sin I\cos\Omega,~\cos I}:$ normal unit vector. It is perpendicular to the satellite's orbital plane
\item[] $M_\odot:$ mass of the Sun
\item[] $\mu_\odot\doteq GM_\odot:$ gravitational parameter of the Sun
\item[] $r_\oplus:$ magnitude of the Earth's position vector with respect to the Sun
\item[] $\epsilon:$ mean obliquity
\item[] $a_\oplus:$  semimajor axis of the heliocentric Earth's orbit
\item[] $n_\textrm{b}^\oplus \doteq \sqrt{\mu_\odot a_\oplus^{-3}}:$  Keplerian mean motion of the heliocentric Earth's orbit
\item[] $P_\oplus\doteq 2\uppi{n_\textrm{b}^\oplus}^{-1}:$ orbital period of the heliocentric Earth's orbit
\item[] $e_\oplus:$  eccentricity of the heliocentric Earth's orbit
\item[] $I_\oplus:$  inclination of the orbital plane of the heliocentric Earth's orbit to the Earth's equator
\item[] $\Omega_\oplus:$  longitude of the ascending node  of the heliocentric Earth's orbit
\item[] $\Omega^0_\oplus:$  initial value of the longitude of the ascending node  of the heliocentric Earth's orbit
\item[] $\dot\Omega_\oplus:$ frequency of the node of the heliocentric Earth's orbit
\item[] $T_{\Omega_\oplus}\doteq 2\uppi \dot\Omega_\oplus^{-1}:$ period of the node of the heliocentric Earth's orbit
\item[] ${\bds L}^\oplus:$ orbital angular momentum per unit mass of the heliocentric Earth's orbit
\item[] $M_\textrm{X}:$ mass of the 3rd body X (Sun $\odot$ or Moon $\leftmoon$)
\item[] $\mu_\textrm{X}\doteq GM_\textrm{X}:$ gravitational parameter of the 3rd body X (Sun $\odot$ or Moon $\leftmoon$)
\item[] $r_\textrm{X}:$ magnitude of the geocentric position vector of the 3rd body X (Sun $\odot$ or Moon $\leftmoon$)
\item[] $a_\textrm{X}:$  semimajor axis of the geocentric orbit of the 3rd body X (Sun $\odot$ or Moon $\leftmoon$)
\item[] $P_\textrm{X}:$ orbital period of the geocentric orbit of the 3rd body X (Sun $\odot$ or Moon $\leftmoon$)
\item[] $e_\textrm{X}:$  eccentricity of the geocentric Earth's orbit of the 3rd body X (Sun $\odot$ or Moon $\leftmoon$)
\item[] $I_\textrm{X}:$  inclination of the orbital plane of the geocentric orbit of the 3rd body X to the Earth's equator (Sun $\odot$ or Moon $\leftmoon$)
\item[] $\Omega_\textrm{X}:$  longitude of the ascending node  of the geocentric orbit of the 3rd body X (Sun $\odot$ or Moon $\leftmoon$)
\item[] $T_{\Omega_{\leftmoon}}:$ period of the node of the geocentric Moon's orbit
\item[] $\uplambda_\odot:$ Sun's ecliptic longitude
\item[] $\hat{s}=\grf{\cos\lambda_\odot,~\sin\lambda_\odot\cos\epsilon,~\sin\lambda_\odot\sin\epsilon}:$ versor of the geocentric Sun's direction
\item[] $i_\odot:$ angle between the geocentric Sun's direction and the satellite's orbital plane
\end{description}
\section{Tables and Figures}\lb{appenb}
\begin{table*}
\caption{Relevant physical and orbital parameters used in the text. Most of the reported values come from \citet{2010ITN....36....1P,2001CeMDA..79..201I,2009P&SS...57.1405D} and references therein. The source for the orbital elements characterizing the heliocentric orbit of the Earth and the geocentric orbit of the  Moon, both referred to the mean Earth's equator at the reference epoch J2000.0,  is the freely consultable database JPL HORIZONS on the Internet at https://ssd.jpl.nasa.gov/?horizons from which they were retrieved by choosing the time of writing this paper as input epoch. For the sake of completeness, we quote also the values of some parameters ($\omega_\oplus,~\omega_{\leftmoon}$) not used to produce the numerical calculation and the plots displayed here. For the level of accuracy with which some of the parameters listed here are currently known, see the main text.
}\lb{tavola0}
\begin{center}
\begin{tabular}{|l|l|l|}
  \hline
Parameter  & Units & Numerical value \\
\hline
$G$ & $\textrm{kg}^{-1}~\textrm{m}^3~\textrm{s}^{-2}$ & $6.67259\times 10^{-11} $\\
$c$ & $\textrm{m~s}^{-1}$ & $2.99792458\times 10^8$\\
$\upmu_0$ & $\textrm{kg~m~A}^{-2}~\textrm{s}^{-2}$ & $1.25664\times 10^{-6}$\\
$\mu_\oplus$ & $\textrm{m}^3~\textrm{s}^{-2}$ & $ 3.986004418\times 10^{14}$ \\
$R_\oplus$ & $\textrm{m}$ & $6.3781366\times 10^{6}$\\
$\textrm{m}_\oplus$ & $\textrm{A~m}^2$ & $7.84\times 10^{22}$\\
${\overline{C}}_{2,0}$ & $-$ & $-4.84165299806\times 10^{-4}$\\
$g_\oplus$ & $\textrm{m~s}^{-2}$ & $9.7803278$\\
$k^{(0)}_{2,1,{K_1}}$ & $-$ & $0.257$\\
$H^1_2\ton{K_1}$ & $\textrm{m}$ & $0.3687012$\\
$\delta_{2,1,{K_1}}$ & $\textrm{deg}$ & $-0.3$\\
$\rho_\textrm{w}$ & $\textrm{kg~m}^{-3}$ & $1.025\times 10^3$ \\
$k^{'}_2$ & $-$ & $-0.3075$\\
$C^{+}_{2,1,{K_1}}$ & $\textrm{m}$ & $0.0283$\\
$\varepsilon^{+}_{2,1,{K_1}}$ & $\textrm{deg}$ & $320.6$\\
$m_\textrm{LAGEOS}$ & $\textrm{kg}$ & $411$\\
$\mu_\odot$ & $\textrm{m}^3~\textrm{s}^{-2}$ & $ 1.32712440018\times 10^{20}$ \\
$\epsilon$ & $\textrm{deg}$ & $23.4393$\\
$a_\oplus$ & $\textrm{au}$ & $0.9992521882390240$\\
$e_\oplus$ & $-$ & $0.01731885059206812$ \\
$I_\oplus$ & $\textrm{deg}$ & $23.43866881079952$ \\
$\Omega_\oplus$ & $\textrm{deg}$ & $359.9979832232821$ \\
$\dot\Omega_\oplus$ & $\textrm{deg~cty}^{-1}$ & $-0.24123856$ \\
$\omega_\oplus$ & $\textrm{deg}$ & $104.4327857096247$ \\
$\mu_{\leftmoon}$ & $\mu_\oplus$ & $1.23000371\times 10^{-2}$ \\
$a_{\leftmoon}$ & $\textrm{km}$ & $385,734$ \\
$e_{\leftmoon}$ & $-$ & $0.05183692147447081$\\
$I_{\leftmoon}$ & $\textrm{deg}$ & $20.79861698590651$ \\
$\Omega_{\leftmoon}$ & $\textrm{deg}$ & $12.09689740287468$\\
$\omega_{\leftmoon}$ & $\textrm{deg}$ & $106.6017252121480$\\
\hline
\end{tabular}
\end{center}
\end{table*}
\end{appendices}
\bibliography{Gclockbib,semimabib,PXbib}{}


\end{document}